\documentclass[twocolumn,aps,floatfix,showpacs,prl]{revtex4-2}

\usepackage{amsmath}
\usepackage{amssymb}
\usepackage{esint}
\usepackage{graphicx}
\usepackage{bbold}
\usepackage[normalem]{ulem}

\usepackage{setspace}

\usepackage{breakurl}
\usepackage{seqsplit}
\usepackage[hidelinks,colorlinks]{hyperref}
\hypersetup{citecolor=[rgb]{0.25,0.14,0.63}}
\hypersetup{urlcolor=[rgb]{0.25,0.14,0.63}}
\hypersetup{linkcolor=black}

\makeatletter

\usepackage{color}

\begin{document}


\title{Benchmarks of Generalized Hydrodynamics for 1D Bose Gases}
\author{R. S. Watson}
\author{S. A. Simmons}
\author{K. V. Kheruntsyan}
\address{School of Mathematics and Physics, University of Queensland, Brisbane, Queensland 4072, Australia}

\date{\today}
\begin{abstract}
Generalized hydrodynamics (GHD) is a recent theoretical approach that is becoming a go-to tool for characterizing out-of-equilibrium phenomena in integrable and near-integrable quantum many-body systems. Here, we benchmark its performance against an array of alternative theoretical methods, for an interacting one-dimensional Bose gas described by the Lieb-Liniger model. In particular, we study various quantum shock wave scenarios, along with a quantum Newton's cradle setup, for various interaction strengths and initial temperatures. We find that GHD generally performs very well at sufficiently high temperatures or strong interactions. For low temperatures and weak interactions, we highlight situations where GHD, while not capturing interference phenomena on short lengthscales, can describe a coarse-grained behaviour based on convolution averaging that mimics finite imaging resolution in ultracold atom experiments. In a quantum Newton's cradle setup based on a double-well to single-well trap quench, we find that GHD with diffusive corrections demonstrates excellent agreement with the predictions of a classical field approach.
\end{abstract}
\maketitle

\emph{\textbf{Introduction.}}---The study of dynamics of integrable and near-integrable quantum many-body systems has been a thriving area of research for more than a decade 
since the landmark experiments on relaxation in the quantum Newton's cradle setup \cite{kinoshita2006quantum} and in coherently split one-dimensional (1D) Bose gases \cite{hofferberth2007non}. During this time, an in-depth understanding of the mechanisms of thermalization and emergent out-of-equilibrium phenomena within these systems has been developed \cite{rigol2007relaxation,rigol2008thermalization,cr10,pssv11,ge16,Langen207}. A recent breakthrough in this area 
has been the discovery of the theory of generalized hydrodynamics (GHD) \cite{Bertini_2016_Transport,Castro-Alvaredo_2016_Emergent} (for recent reviews, see \cite{doyon2019lecture,bouchoule2022generalized,essler2022short}). This new theory is capable of simulating large-scale dynamics of integrable and near-integrable systems across a significantly broader range of particle numbers and interaction strengths than those accessible using previous approaches \cite{largescale_Doyon,GHD_onatomchip,malvania2020generalized}. Because of its broad applicability, GHD is currently regarded as well on its way to becoming ``a standard tool in the description of strongly interacting 1D quantum dynamics close to integrable points" \cite{malvania2020generalized}.

In the years since its discovery, GHD has been rapidly developed to include diffusive terms \cite{DeNardis_Diffusion_2018,gopalakrishnan2018hydrodynamics,bastianello_thermalization_2020,durnin2021diffusive,bastianello2021hydrodynamics,Bulchandani_2021}, particle loss \cite{Bouchoule_AtomLoss_2020},
calculations of quantum and Euler-scale correlations \cite{ruggiero2020quantum,ruggiero2021quantum,Doyon_correlation_2018,moller2020euler,De_Nardis_2022,Alba_2021}, as well as the incorporation of numerous beyond-Euler scale effects \cite{Fagotti_higherorder_2017,Panfil_Linearized_2019,moller2020extension,Bastianello_dephasing_2020} (see also \cite{Bu_a_2021,Borsi_2021,El_2021,Cubero_2021} in a special issue). Recently, GHD applied to a 1D Bose gas has been experimentally verified in a variant of the quantum Newton's cradle setup in the weakly interacting regime \cite{GHD_onatomchip}, and in a harmonic trap quench in the strongly interacting regime \cite{malvania2020generalized}. In both cases, GHD provided an accurate coarse-grained model of the dynamics, exceeding conventional (classical) hydrodynamics. In addition to comparisons with experiments, GHD was benchmarked against other established theoretical approaches---most prominently for the 1D Bose gas and $XXZ$ spin chain \cite{Castro-Alvaredo_2016_Emergent,Bertini_2016_Transport,largescale_Doyon,bastianello2019generalized,moller2020euler,GHD_onatomchip,Bastianello_dephasing_2020,ruggiero2020quantum,UniversalShock,bulchandani2018bethe,urichuk2019spinDrude,doyon2018geometric}. As the purpose of these initial benchmarks was to validate GHD, the typical dynamical scenarios considered were in regimes where GHD was expected to be a valid theory. In all such cases GHD demonstrated very good agreement with the alternative approaches. On the other hand, in scenarios involving, for example, short wavelength density oscillations due to interference phenomena (which are not captured by GHD), it was conjectured that GHD would nevertheless adequately describe spatial coarse-grained averages of the more accurate theories \cite{largescale_Doyon,GHD_onatomchip,moller2020extension}. More generally, it is of significant interest to scrutinize the performance of GHD by extending its benchmarks to a more challenging set of dynamical scenarios. This is important for understanding exactly how GHD breaks down when it is pushed towards and beyond the limits of its applicability.

In this Letter, we systematically benchmark the performance of GHD for the 1D Bose gas in several paradigmatic out-of-equilibrium scenarios. In particular, we focus on the regime of dispersive quantum shock waves emanating from a localized density bump of the type explored recently in Ref.~\cite{whatisqushock}. We use an array of theoretical approaches, including finite temperature $c$-field methods, the truncated Wigner approximation, and the numerically exact infinite matrix product state (iMPS) method, spanning the entire range of interaction strengths, from the nearly ideal Bose gas to the strongly interacting Tonks-Girardeau (TG) regime. We also analyse the dynamics of a localized density dip which sheds grey solitons, hence benchmarking GHD in scenarios not previously considered. In doing so we address the question of how well GHD predictions agree with coarse-grained averaging of the results of the more accurate theoretical approaches. Additionally, we explore the dynamics of a thermal quasicondensate in a quantum Newton's cradle setup \cite{kieran_paper,GHD_onatomchip,GHD_newtonscradle} using Navier-Stokes type diffusive GHD \cite{de2019diffusion,DeNardis_Diffusion_2018,durnin2021diffusive}, and address the question of characteristic thermalization rates \cite{kieran_paper,bastianello_thermalization_2020}.

\emph{\textbf{Expansion from a localized density bump.}}---We begin our analysis by considering dispersive quantum shock waves of the type studied recently in Ref.~\cite{whatisqushock}. More specifically, we first focus on the weakly interacting regime of the 1D Bose gas of $N$ particles, and consider the dynamics of the oscillatory shock wave train generated through a trap quench from an initially localized perturbation on top of a flat background to free propagation in a uniform box of length $L$ with periodic boundary conditions \cite{Damski_2004,Damski_2006}. The weakly interacting regime is characterized by the Lieb-Liniger \cite{liebliniger,kheruntsyan2005finite} dimensionless interaction parameter $\gamma_\mathrm{bg}=mg/\hbar^2\rho_\mathrm{bg}\ll1$, defined with respect to the background particle number density, $\rho_\mathrm{bg}$, where $g>0$ is the strength of repulsive contact interaction and $m$ is the mass of the particles.

\begin{figure}[t!]                             
   \includegraphics[width=8.3cm]{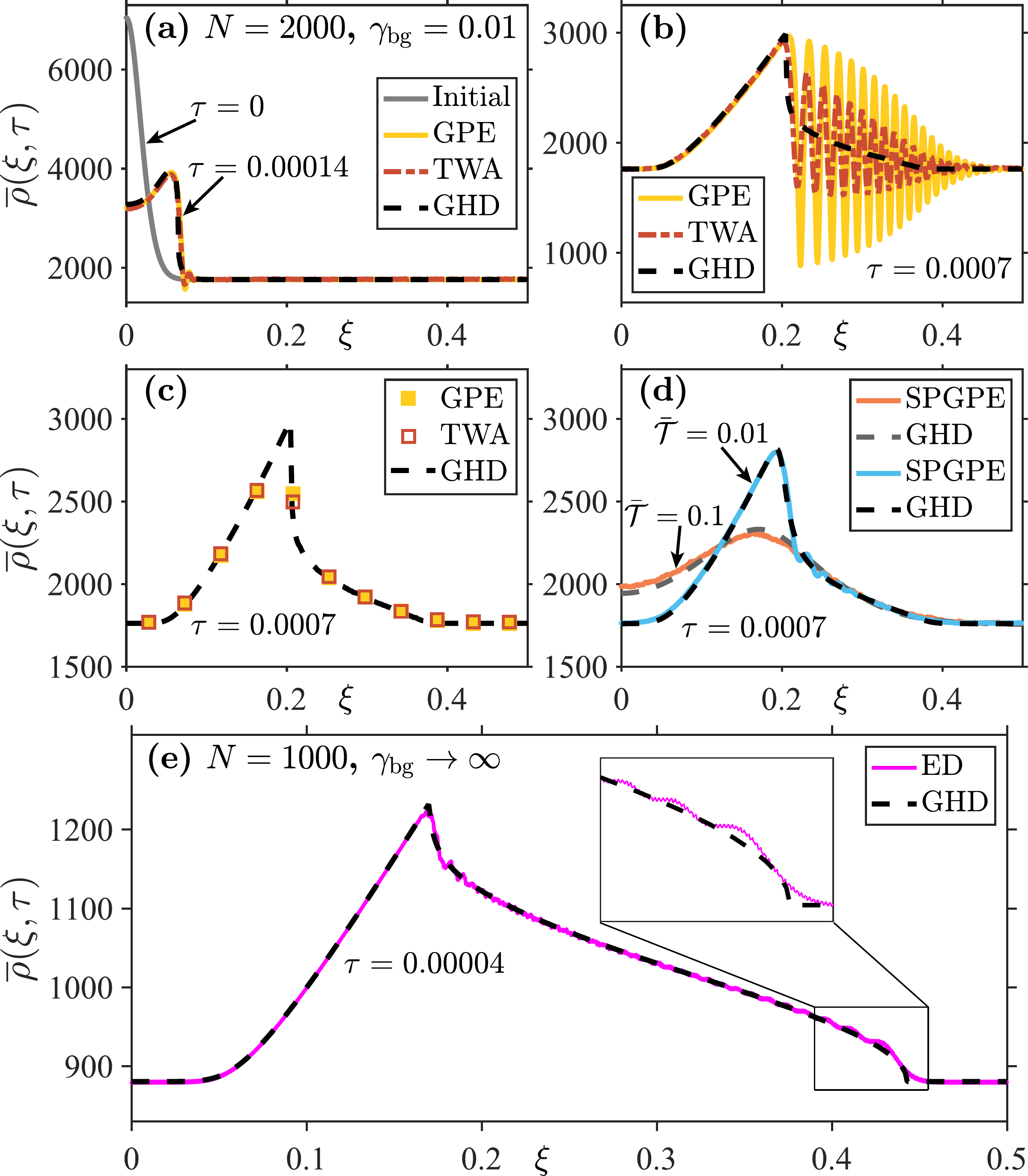}
   \caption{Dimensionless density profiles $\overline{\rho}=\rho L$ of quantum shock waves in the 1D Bose gas, as a function of the dimensionless coordinate $\xi \!\equiv\! x/L$ at different times $\tau \!\equiv\! \hbar t / m L^2$. In (a) we show the initial ($\tau\!=\!0$) and time-evolved ($\tau\!=\!0.00014$)  profiles of a weakly interacting gas    at zero temperature, for $\gamma_\mathrm{bg}\!=\! 0.01$ and $N\!=\!2000$ (with $N_\mathrm{bg}\! \simeq \!1761$ the number of particle in the background). Due to the symmetry about the origin, we only show the densities for $\xi\!>\!0$. In (b), the time-evolved profile is shown at $\tau \!=\! 0.0007$. Panel (c) demonstrates the results of finite resolution averaging    of both GPE and TWA data from (b) and compares them with the same GHD result. Panel (d) shows the same system as in (b), but at finite temperatures, simulated using the stochastic projected GPE (SPGPE) \cite{whatisqushock}; the dimensionless temperature $\overline{\mathcal{T}}$ here is defined according to $\overline{\mathcal{T}} \!=\! T / T_d$, where $T_d \!= \!\hbar^2 \rho_\mathrm{bg}^2 / 2 m k_B$ \cite{kheruntsyan2005finite}. Panel (e) compares GHD predictions with exact diagonalization (ED) results in the TG regime ($\gamma_\mathrm{bg} \!\to\! \infty$) for $N\!=\!1000$ ($N_\mathrm{bg} \!\simeq \!884$), at $\tau=0.00004$. In all examples, the initial profiles are characterized by the amplitude height $\beta\!=\!1$ and dimensionless width of the bump $\overline{\sigma}\!=\!0.02$; see Appendix A for details.    } 
  \label{fig:Bump_largeN}
\end{figure}

In our first example, we consider the case of a large total number of particles, $N\!=\!2000$, and $\gamma_\mathrm{bg}\!=\!0.01$, so that the gas is in the Thomas-Fermi regime where the interaction energy per particle dominates the kinetic energy. We assume that the gas is initialized in the zero-temperature ($T\!=\!0$) ground state of a dimple trap that results in the density profile of Eq.~\eqref{eq:initialdensity} given in Appendix A. At time $\tau\!=\!0$, the dimple trap is suddenly switched off, and we follow the evolution of the system in a uniform 1D trap. In Figs.~\ref{fig:Bump_largeN}\,(a) and (b), we show snapshots of the density profiles at different times, and compare the GHD results with those obtained using the mean field Gross-Pitaevskii equation (GPE) and the truncated Wigner approximation (TWA) which incorporates the effect of quantum fluctuations ignored in the GPE \cite{cfield}. The snapshot at $\tau\!=\!0.00014$, which corresponds to the onset of a shock formation due to a large density gradient, shows excellent agreement between GHD and the more accurate microscopic approaches. Such an agreement  at early times is remarkable given that GHD, which is derived here at Euler scale \cite{spohn2012large}, becomes formally exact only in the limit of infinitely large length and time scales \cite{largescale_Doyon,doyon2019lecture,moller2020euler}.

Past this time, the GPE and TWA show the formation of an oscillatory shock wave train, which has been identified in Ref.~\cite{whatisqushock} as a result of self-interference of the expanding density bump with its own background.
The interference contrast in this regime is generally large, even though the quantum fluctuations present in the TWA approach cause a visible reduction in contrast compared with the mean-field GPE result. The GHD prediction, on the other hand, completely fails to capture the oscillations, as these occur on a microscopic lengthscale. The characteristic period of oscillations here (which we note are chirped) 
is given approximately by the healing length $l_h\!=\!\hbar/\sqrt{mg\rho_\mathrm{bg}}$ ($l_h/L = 0.0057$) which is smaller than the width $\sigma$ ($\sigma/L=0.02$) of the initial bump and hence represents the shortest lengthscale of the problem in the bulk of the shock wave train.
Thus, even though the local density approximation (required for GHD to be applicable to an inhomogeneous system in the first place) is valid for the initial Thomas-Fermi density profile, the failure of GHD at later times is expected since it is not supposed to capture phenomena on microscopic lengthscales, which emerge here dynamically.

\begin{figure*}[tp]
    \includegraphics[width=17.4cm]{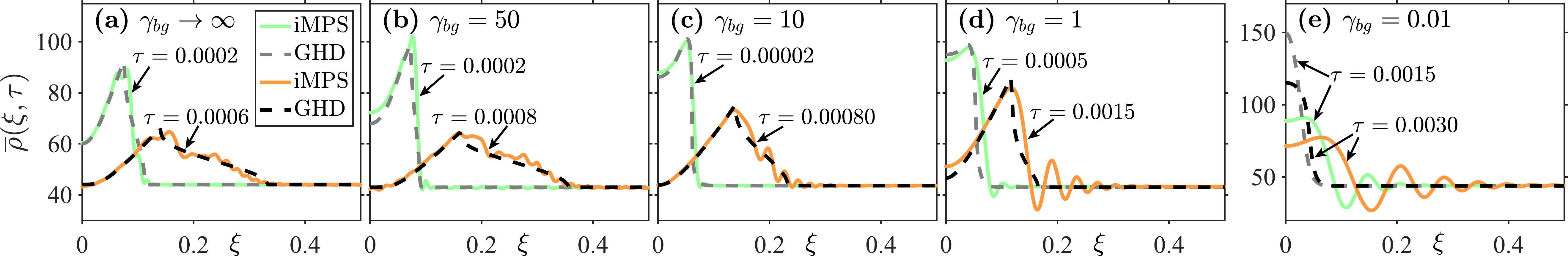}
    \caption{Quantum shock waves at zero temperature for $N=50$ particles ($N_\mathrm{bg} \simeq 44.03$), over the entire range of interaction strengths. In all examples, the initial density profiles (not shown) closely match Eq.~\eqref{eq:initialdensity} in Appendix A, with $\beta=1$ and $\overline{\sigma}=0.02$. In all panels, we show the GHD (dashed lines) and iMPS (full lines) results for the evolved density profiles at two time instances. In (a) there is no phase coherence beyond the mean interparticle separation ($1/\rho_{bg} L \simeq 0.0227$), whereas in (e) the shortest lengthscale that determines the characteristic period of oscillations is given by the width of the initial Gaussian bump $\sigma$ ($\sigma/L=0.02$), which is much smaller than the healing length $l_h$ ($l_h/L=0.227$).}
\label{fig:Bump_lowN}
\end{figure*}

Despite this failure, GHD clearly captures the average density of the oscillations for the fully formed shock wave train, similar to that shown in \cite{essler2022short}. This is consistent with the analysis of Bettelheim \cite{bettelheim2020whitham}, who showed that the Whitham approach, which allows one to write equations for averaged quantities in the oscillatory shock wave train, is equivalent to GHD in the semiclassical limit ($c \!=\!mg/\hbar^2\!\to\!0$) of the Lieb-Liniger model \cite{whithamBook,kamchatnov2000nonlinear}. This is also consistent with the expectation that GHD in an interfering region would correspond to a coarse-grained average density \cite{GHD_onatomchip,largescale_Doyon}.  To quantitatively assess this expectation, we perform a type of convolution averaging that mimics the finite resolution of \emph{in-situ} imaging systems used in quantum gas experiments (see Appendix B). As the imaging resolution is usually unable to resolve wavelengths on the order of the healing length (typically in the submicron range), one expects that such averaging will smear out the interference fringes seen in the GPE and TWA data---just as GHD implicitly does. In Fig.~\ref{fig:Bump_largeN}(c) we show the results of convolution-averaged density profiles performed on the GPE and TWA data of Fig.~\ref{fig:Bump_largeN}(b) and compare them with the same GHD curve. The level of agreement between all three curves is now remarkable---a result which was not \textit{a priori} obvious for both GPE and TWA under this model of coarse-graining. This highlights the quantitative success of GHD in describing the dynamics on large scale despite interference or short-wavelength phenomena being present. 

In our second set of examples, shown in Fig.~\ref{fig:Bump_largeN}(d), we consider the same shock wave scenario, except now for a phase fluctuating quasicondensate at finite temperatures. Here, the effect of thermal fluctuations is expected to lead to a smearing of the interference contrast due to a reduced thermal phase coherence length in the system, $l_T\!=\!\hbar^2\rho_{\mathrm{bg}}/mk_BT$ \cite{Mora-Castin-2003,Cazalilla_2004,Bouchoule-Arzamasovs-2012}. A well-established theoretical approach to model this is a $c$-field stochastic projected GPE (SPGPE) approach \cite{castin2000,Blakie_cfield_2008} (see also \cite{kieran_paper,bouchoule2016,bayocboc2021dynamics,bayocboc2022frequency}), and we indeed observe such smearing in Fig.~\ref{fig:Bump_largeN}(d) \cite{Thermal_Coherence_length}, in addition to seeing the expected very good agreement of GHD with these $c$-field results.

Our third example is shown in Fig.~\ref{fig:Bump_largeN}(e) and lies in the TG regime of infinitely strong interactions, $\gamma_\mathrm{bg}\!\to \!\infty$. It further illustrates the same observation---that the performance of GHD improves with the loss of phase coherence in the system, wherein interference phenomena are suppressed. Here, we compare the predictions of GHD for the shock wave scenario at $T\!=\!0$ with the results of exact diagonalization. In the TG regime, the system does not posses phase coherence beyond the mean interparticle separation $1/\rho_{bg}$, hence the absence of interference fringes in the evolution of a density bump whose initial width is larger than $1/\rho_{bg}$ \cite{whatisqushock}. Accordingly, we see very good agreement of GHD with exact diagonalization, ignoring the small-amplitude density ripples that can be seen in the exact result. Such density ripples (which we note have different origin to Friedel oscillations) have been predicted to occur in the ideal Fermi gas by Bettelheim and Glazman~\cite{quripples} (see also \cite{Friedel_vs_Glazman}). By the Fermi-Bose mapping \cite{Girardeau_1960,Girardeau-Wright-2000}, these same ripples should emerge in the TG gas, which we confirm here through exact diagonalization. However, their description lies beyond the scope of GHD as a large-scale theory \cite{GHD_and_Wigner}.

The final set of examples for the evolution of a density bump is shown in Fig.~\ref{fig:Bump_lowN}. Here, we consider a range of interaction strengths, starting from very strong and going back [from (a) to (e)] to weak interactions, all at zero temperature and $N\!=\!50$. We compare the GHD results with iMPS simulations, which are numerically exact at all interaction strengths \cite{whatisqushock}. At this relatively low particle number, the strongly interacting regime displays Friedel oscillations which appear in the iMPS result and are, as expected, absent from the prediction of GHD. However, there is generally good agreement between GHD and iMPS at large scale. As the interaction strength is reduced, and hence the phase coherence of the gas increases,
the Friedel oscillations disappear and interference fringes return, which now have period $\sim\!\sigma$ (with $\!\sigma<\!l_h$) since the gas is no longer in the Thomas-Fermi regime. The worst performance of GHD is observed for $\gamma_{\mathrm{bg}}\!=\!0.01$, which lies in the nearly ideal (noninteracting) Bose gas regime for $N\!=\!50$. In this regime, the local density approximation, intrinsic to GHD \cite{largescale_Doyon,malvania2020generalized,GHD_onatomchip,kheruntsyan2005finite}, is no longer valid even for the initial density profile, and we see that Euler-scale GHD breaks down both spatially and temporally, explaining the failure of GHD to agree with iMPS results even in the coarse-grained sense.

In addition to considering the dynamics of a localized density bump, we have also analyzed evolution of an initial density dip in a uniform background. This scenario is known to shed a train of grey solitons in the mean-field GPE treatment  \cite{Damski_2004,Damski_2006,Engels_Hoefer_2008}, and the results of comparison of GHD simulations with those of GPE and TWA are presented in Appendix C. The overall conclusions regarding the performance of GHD in this scenario are similar to those for a density bump, including good agreement of GHD with coarse-grained averages of GPA and TWA results in the soliton train region.

\emph{\textbf{Quantum Newton's cradle in a thermal quasicondensate.}}---Our final scenario for benchmarking GHD is in a variant of the quantum Newton's cradle setup for a weakly interacting 1D Bose gas in the quasicondensate regime. Namely, we analyze the release from a symmetric double-well trap to a single-well harmonic trap of frequency $\omega$, similar to the type utilized  in Ref.~\cite{GHD_onatomchip}. Here, we use the SPGPE to simulate collisional dynamics and eventual thermalization, as in Ref.~\cite{kieran_paper}, and for the sake of one-to-one comparison, we also simulate the same system using the Navier-Stokes type of diffusive GHD \cite{DeNardis_Diffusion_2018,de2019diffusion}, solved using a second-order backwards-implicit algorithm 
\cite{gopalakrishnan2018hydrodynamics,bastianello2019generalized,moller2020introducing}.

\begin{figure}[tbp]
\hspace*{-0.3cm} 
   \includegraphics[width=8.4cm]{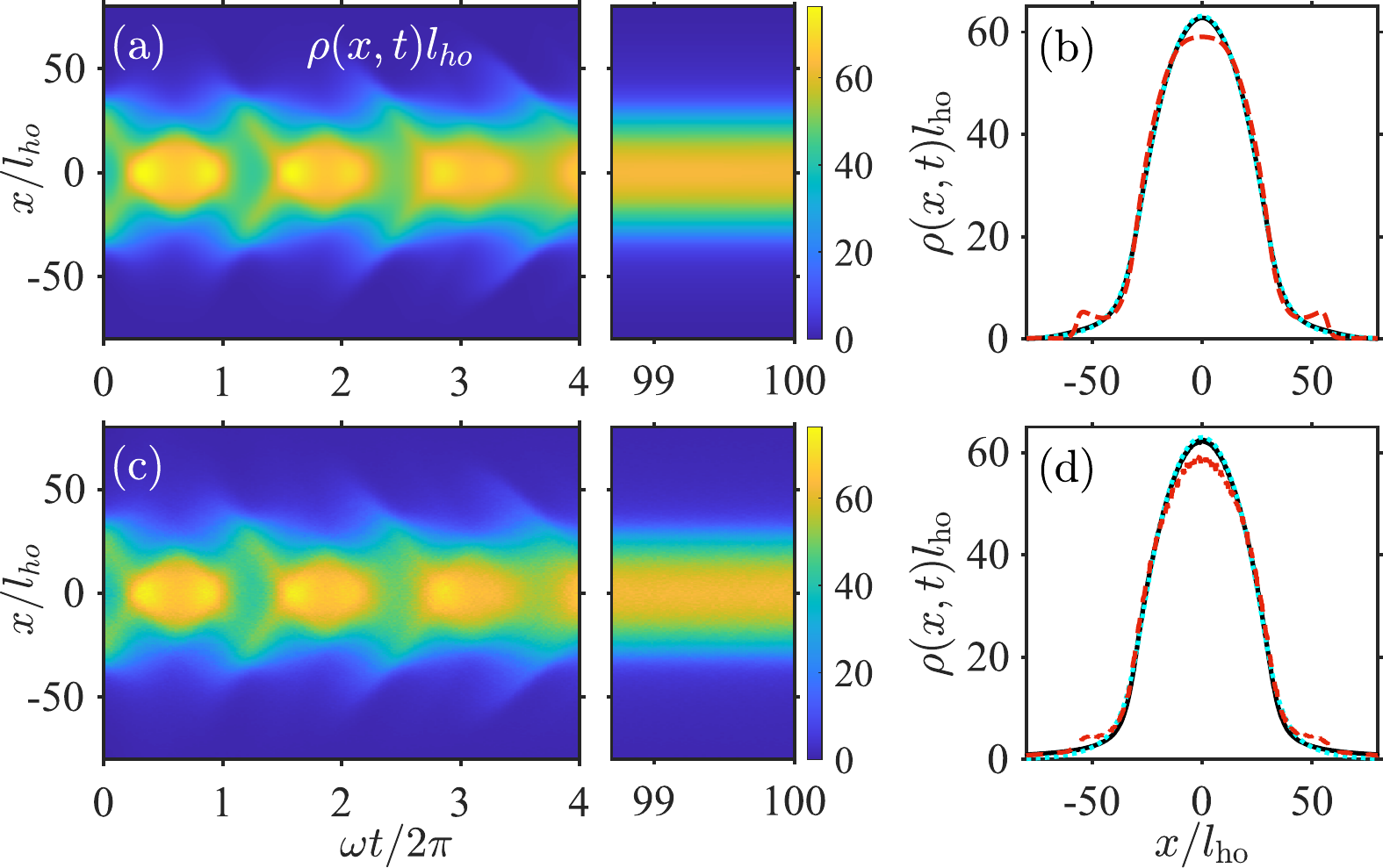}   
   \caption{Evolution and thermalization of the density distribution $\rho(x,t)$ in a quantum Newton's cradle setup initialized from a double-well to single-well trap quench, simulated using (a)-(b) Navier-Stokes GHD, and (c)-(d) SPGPE. The initial cloud of $N=3340$ atoms at temperature $\widetilde{T} = 205$ (in harmonic oscillator units) is prepared in a thermal equilibrium state of a symmetric double-well trap potential (see Appendix D for details). Panel (b) demonstrates the relaxed density profile of the Navier-Stokes GHD evolution at $t\!=\!100/(\omega/2\pi)$ (black solid line), alongside a best-fit thermal equilibrium profile from Yang-Yang thermodynamics at $\widetilde{T}\!\simeq \!213$ (cyan dotted line), and an additional GHD density profile at earlier time $t\!=\!6.79/(\omega/2\pi)$ (red dashed line). Panel (d) is the same as (b), but for the SPGPE, with the relaxed density profile at $t\!\simeq\!100/(\omega/2\pi)$, Yang-Yang thermodyanmic density profile of $\widetilde{T} \!\simeq \!216$, and an additional density profile at $t\!=\!6.81/(\omega/2\pi)$.
   }
  \label{fig:NewtonsCradle}
\end{figure}

Comparison of the results using the two methods are shown in Fig.~\ref{fig:NewtonsCradle}, where we illustrate the evolution of the density distribution [(a) --  for diffusive GHD, and (c) -- for SPGPE] over the initial few oscillations, as well as after sufficiently long time, when the system has already thermalized. In Ref.~\cite{SupMat} we give further details of how the final relaxed states were assessed within GHD and the SPGPE, whereas here, in Figs.~\ref{fig:NewtonsCradle}\,(b) and (d), we simply show the respective relaxed density profiles, along with their corresponding thermal equilibrium profiles from Yang-Yang thermodynamics \cite{yang1969thermodynamics,Karen_Yang_2008, kheruntsyan2005finite,SupMat}, as well as density profiles at earlier times illustrating their contrast to the relaxed state. The overall conclusion here is that GHD demonstrates excellent agreement with SPGPE  in both short- and long-term dynamics, as well as in the characteristic thermalization rate \cite{GHD_time_window}.

We have also simulated the quantum Newton's cradle experiment in the original Bragg pulse scenario \cite{kinoshita2006quantum}, except in a weakly interacting quasicondensate regime. In this scenario, we observe different thermalization rates in GHD and SPGPE simulations, and we discuss these results and the reasons behind the discrepancy in the Supplementary Material \cite{SupMat}.

\emph{\textbf{Summary.}}---
We have benchmarked GHD in a variety of out-of-equilibrium scenarios in a 1D Bose gas against alternative theoretical approaches which are not limited to long-wavelength excitations. In particular, we have focused on systems supporting dispersive quantum shock waves and soliton trains, demonstrating that GHD generally agrees with the predictions of these approaches at sufficiently high temperatures and strong interactions. Here, the good agreement stems from a reduced phase coherence length of the gas, which in turn leads to a suppression of interference phenomena and therefore an absence of high-contrast short-wavelength interference fringes in the density. At low temperatures and weak interactions, where interference phenomena are more pronounced, the predictions of GHD only agree with a coarse-grained convolution averaging approximation. The effect of such averaging is similar to having finite imaging resolution in quantum gas experiments, and explains why GHD may perform well when compared to experiments, whilst departing from the predictions of theoretical approaches that are valid at short wavelengths. We have also benchmarked Navier-Stokes GHD within a quantum Newton's cradle setup for a double-well to single-well trap quench of a weakly interacting quasicondensate, observing excellent agreement with the SPGPE in both transient dynamics and final relaxed state, as well as in the characteristic relaxation timescale.

K.\,V.\,K. acknowledges stimulating discussions with I.~Bouchoule, M.~J.~Davis, and D.~M.~Gangardt. This work was supported through Australian Research Council (ARC) Discovery Project Grants No. DP190101515.

~

\emph{\textbf{Appendix A: Parametrization of the density bump.}}---The initial density profile in Fig.~\ref{fig:Bump_largeN}\,(a), in dimensionless units, is set to 
\begin{equation}\label{eq:initialdensity}
   \overline{\rho}(\xi,\tau\!= \!0) \!=\! \overline{\rho}_\mathrm{bg} \big( 1 \!+\! \beta e^{-\xi^2/2 \overline{\sigma}^2} \big)^2,
\end{equation}
where the dimensionless coordinate, time, and density are introduced, respectively, according to $\xi \!\equiv\! x/L$, $\tau \!\equiv\! \hbar t / m L^2$, and $\overline{\rho}(\xi,\tau)\!\equiv\!\rho(x,t)L$, with $\overline{\rho}_\mathrm{bg}\!=\!\rho_\mathrm{bg}L\!=\!N_\mathrm{bg}$ being the dimensionless background density equivalent to the total number of particles in the background, $N_\mathrm{bg}\!=\!N/\big( 1+ \frac{\sqrt{\pi} \beta \sigma}{L} [ \beta\, \mathrm{erf}(\frac{L}{2\sigma}) +2 \sqrt{2}  \,\mathrm{erf} (\frac{L}{2\sqrt{2}\sigma}   )  ] \big)$ from the normalization. In addition, the width and amplitude of the bump above the background are characterized by the dimensionless parameters $\bar{\sigma}\equiv\sigma/L$ and $\beta>0$, respectively.

The associated trapping potential that is required for preparation of such a density profile as an initial ground or thermal equilibrium state of the 1D Bose gas in different regimes is discussed in Ref.~\cite{whatisqushock}. Within the mean-field approximation, described by the Gross-Pitaevskii equation, the density profile of Eq. (\ref{eq:initialdensity}) corresponds to the mean field amplitude being initialized as a simple Gaussian bump superimposed on a constant background, $\overline{\Psi}(\xi,\tau\!= \!0) \!=\! \overline{\Psi}_\mathrm{bg} \big( 1 \!+\! \beta e^{-\xi^2/2 \overline{\sigma}^2}\big)$, with $\overline{\rho}_\mathrm{bg}=|\overline{\Psi}_\mathrm{bg}|^2$.

\emph{\textbf{Appendix B: Finite resolution averaging}}---
Finite resolution averaging procedure implemented in Fig.~1(c) 
emulates the finite spatial resolution of experimental absorption imaging systems. Following 
Ref.~\cite{Armijo_ThreeBody_2010}, we denote the impulse response function of the imaging system by $\mathcal{A}(x)$, which we here assume to be a normalized Gaussian. 
The impulse response for a pixel of width $\Delta$ centered at $x_p$ is then,
\begin{equation}
    \mathcal{F}(x) = \int_{x_p - \Delta/2}^{x_p+\Delta/2} dx' \mathcal{A}(x'-x).
\end{equation}
The \emph{measured} atom number in the given pixel is then given by
\begin{equation}
    N_m = \mathcal{N} \int_{-\infty}^{+\infty} dx \mathcal{F}(x) \rho(x),
\end{equation}
where $\mathcal{N}$ provides the correct normalization for the total particle number in the limit of zero pixel width.

In our particular example of such averaging, the density profile $\rho(x)$ (at any given time step, with the time argument $t$ being omitted here for notational simplicity) is convoluted with a Gaussian resolution function of width $w = 1$ $\mu$m and then averaged over a finite pixel size $\Delta= 4.5$ $\mu$m, as in Ref.~ \cite{Armijo_ThreeBody_2010}. These absolute values translate to dimensionless values of $w/L = 0.01$ and $\Delta/L = 0.045$, assuming $L\sim 100$~$\mu$m, with results being generally insensitive to the exact values of these parameters around these typical values.
For comparison, the healing length in this example is equal to $l_h/L\!=\!0.0057$. Considering $^{87}$Rb atoms, which have a scattering length of $a\simeq 5.3$~nm, in a system of size $L=100$~$\mu$m, this corresponds to an absolute healing length of $l_h = 0.57$~$\mu$m. These choices of dimensionless parameters, and $\gamma_{\mathrm{bg}}=0.01$, can be realized at a background density of $\rho_\mathrm{bg} \simeq 1.8 \times 10^7$~m$^{-1}$, with an interaction parameter $g\simeq 2 \hbar \omega_\perp a \simeq 1.4\times 10^{-38} \mathrm{J} \! \cdot \! \mathrm{m}$ \cite{Olshanii1:998}, where $\omega_\perp / 2 \pi \simeq 1.9$~kHz is the frequency of the transverse harmonic trapping potential.

\begin{figure}[tbp]                          
   \includegraphics[width=8.4cm]{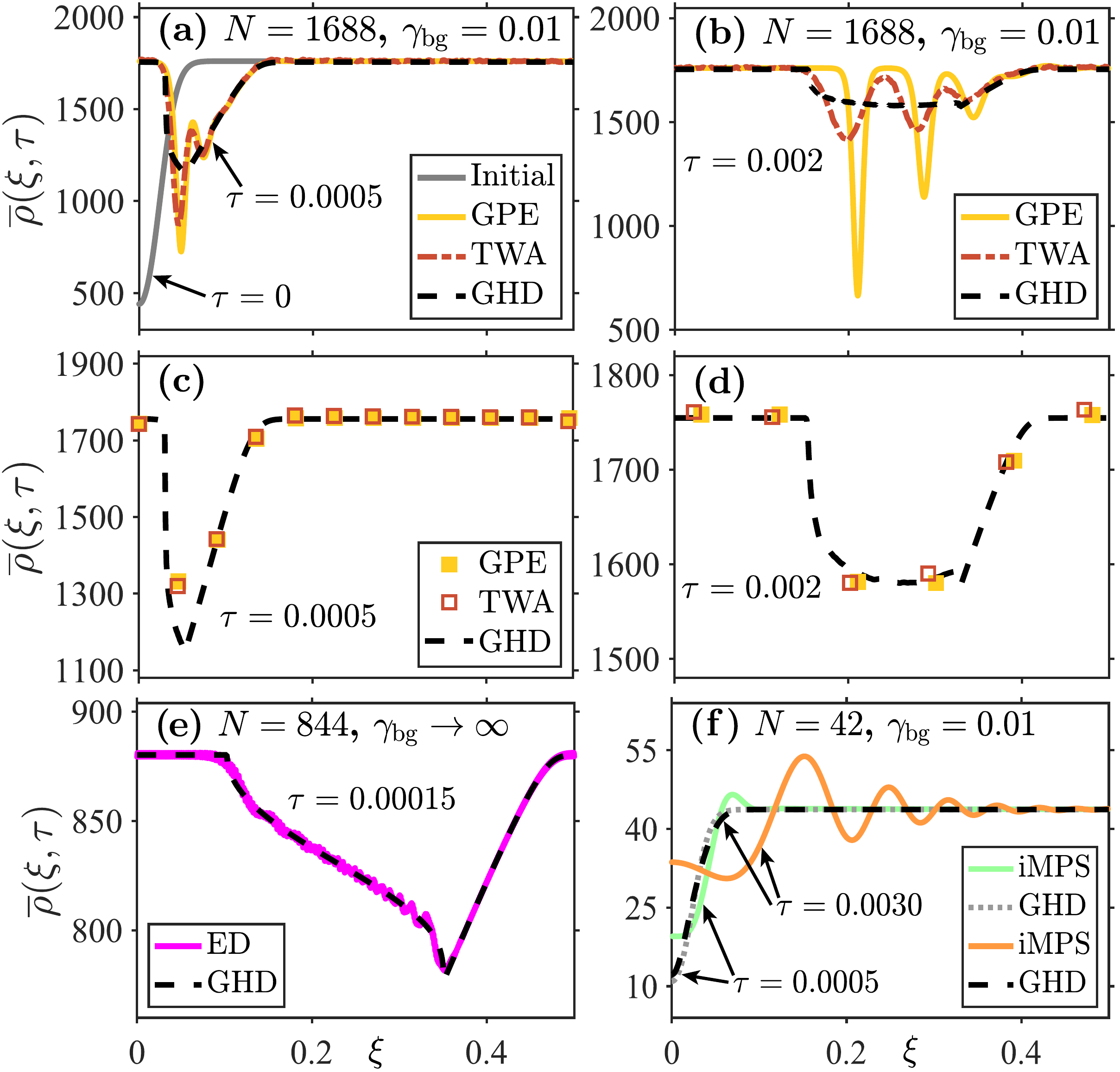}
   \caption{Evolution of a density dip in a 1D Bose gas. Panel (a) shows the initial ($\tau=0$) and time-evolved ($\tau=0.0005$) density profiles from GPE, TWA and GHD simulations, for $\gamma_\mathrm{bg} \!= \!0.01$ and $N \!= \!1688$ ($N_\mathrm{bg} \!\simeq \!1761$); panel (b) shows a time-evolved density profile at a later time ($\tau=0.002$), where we can see a fully formed train of three grey solitons in the mean-field GPE (full yellow) curve.
   Panels (c) and (d) compare the same GHD results (notice the different scale of vertical axis) at $\tau=0.0005$ and $\tau=0.002$ with the outcomes of finite resolution averaging of both GPE and TWA curves. In panel (e), we show a time-evolved snapshot of the density profile in the TG regime ($\gamma_\mathrm{bg} \!\to  \! \infty$) for $N \!= \!844$ ($N_\mathrm{bg} \!\simeq \!880.5$), and compare the GHD result with that of exact diagonalization (ED). Panel (f) is in the nearly ideal Bose gas regime, with $\gamma_\mathrm{bg} \!= \!0.01$, $N \!= \!42$ ($N_\mathrm{bg} \!\simeq \!44$). In all examples, the initial density profile is given by Eq.~\eqref{eq:initialdensity} with $\beta \!=\! -0.5$ and $\overline{\sigma}\!=\!0.02$.} 
  \label{fig:Dip}
\end{figure}

\emph{\textbf{Appendix C: Dynamics of a localized density dip.}}---In this Appendix, we present the results of evolution of a localized density depression, after quenching (at time $\tau\!=\!0$) the initial trap potential with a localized barrier to uniform. We assume that the initial density profile is given by the same Eq.~\eqref{eq:initialdensity}, except with $\beta$ being negative and satisfying $-1 \!<\! \beta\! <\! 0$. 

In Figs.~\ref{fig:Dip}\,(a) and (b), we consider the weakly interacting regime (with $\gamma_\mathrm{bg} \!= \!0.01$) and show the results of the GPE, TWA, and GHD simulations, for a gas with $N\!=\!1688$ atoms and the same $N_\mathrm{bg} \!\simeq \!1761$ as in Fig.~\ref{fig:Bump_largeN}\,(a). In this scenario, the steep gradient of the shock front forms as the background fluid flows inward and tries to fill the density depression. As a result, one first observes the emergence of large-amplitude structures, forming multiple density troughs, which then evolve into a train of grey solitons propagating away from the origin \cite{Damski_2004,Damski_2006,Engels_Hoefer_2008,el_DSW,Gurevich_Pitaevskii,kamchatnov2000nonlinear}. The differences between the TWA and pure mean-field GPE results, seen in Figs.~\ref{fig:Dip}\,(b), are consistent with previous observations \cite{Dziarmaga_2003,Dziarmaga_2004,Ruostekoski_2010} that quantum fluctuations lower the mean soliton speed and fill in the soliton core. The GHD result, on the other hand, fails to capture the solitonic structures, whose characteristic width (on the order of the microscopic healing length) lies beyond the intended range of applicability of GHD. 

However, GHD still manages to adequately capture the coarse-grained description of the density across the soliton train, which is rather remarkable. This is seen in Fig.~\ref{fig:Dip}\,(c) and (d), where we demonstrate the outcomes of finite resolution averaging applied to GPE and TWA results of panels (a) and (b), respectively. Similarly to Fig.~1(c), here we used the same normalized Gaussian resolution function of width $1$\,$\mu$m and adopted $^{87}$Rb atoms as an example species for the relevant parameter values (see Appendix B). For panel (c) we used the same pixel size ($\Delta = 4.5$\,$\mu$m) as before, whereas for panel (d), due to the presence of fully formed grey solitons whose width is on the order of $(2-4)l_h$, we used a twice larger pixel size ($\Delta = 9.0$\,$\mu$m). A larger pixel size here results in $\Delta/l_h\simeq 16\gg 1$, which is required in order to comply with the large-scale framework of GHD. 
 
The last two examples, shown in Figs.~\ref{fig:Dip}\,(c) and (d), correspond, respectively, to the strongly interacting TG and nearly ideal Bose gas regimes. The overall behaviour and conclusions about the performance of GHD in these examples are the same as in the equivalent scenario of the density bump discussed earlier in Figs.~\ref{fig:Bump_largeN}\,(e) and \ref{fig:Bump_lowN}\,(a).

\emph{\textbf{Appendix D: Parametrization of the double-well trap}}---The initial (pre-quench) double-well trap potential is set to { $\widetilde{V}(\widetilde{x}) \!\simeq\! 2.16 \! \times \! 10^{-3} \widetilde{x}^4 \!-\! 5.27 \! \times \! 10^{-1} \widetilde{x}^2$ in dimensionless form, where $\widetilde{x} \!= \!x/l_\mathrm{ho}$, $\widetilde{V} \!=\! V / \hbar \omega$, and $l_\mathrm{ho} \!=\! \sqrt{\hbar/m \omega}$, where $\omega$ is the post-quench single-well harmonic trap frequency. The initial dimensional temperature of the cloud, in harmonic oscillator units, is set to  $\widetilde{T} \!=\! T/(\hbar \omega/k_B)\!\simeq\! 205$. In this configuration, the initial density profile for a total of $N=3340$ atoms is double peaked, with the dimensionless interaction strength at either of the peaks given by $\gamma_{\max}\simeq0.0138$.


\begin{thebibliography}{85}
\bibitem [{The()}]{Thermal_Coherence_length}%
  \BibitemOpen
  \href@noop {} {}\bibinfo {howpublished} {For the examples considered in
  Fig.~1(d), the thermal phase coherence lengths are $l_T/L\!\simeq\!0.1$ for
  $\overline{\mathcal{T}}\!=\!0.01$ and $l_T/L\!\simeq\!0.01$ for
  $\overline{\mathcal{T}}\!=\!0.1$. These values, in turn, are comparable or
  smaller than the width of the initial density bump $\overline{\sigma}=0.02$,
  implying a reduced phase coherence over the extent of the bump and hence loss
  of interference contrast upon expansion of the bump into the background
  \cite{whatisqushock}.}\BibitemShut {Stop}%
\bibitem [{Fri()}]{Friedel_vs_Glazman}%
  \BibitemOpen
  \href@noop {} {}\bibinfo {howpublished} {In the example of
  Fig.~\ref{fig:Bump_largeN}\,(e), which is for $N=1000$ particles, we have
  been able to discriminate between the Friedel oscillations, which have a
  period of 1/1000, and the Bettelheim-Glazman density ripples which have a
  larger oscillation period. This was not possible in the prior work of
  Ref.~\cite{whatisqushock}, which treated a much smaller number of particles
  ($N=50$) in the Tonks-Girardeau regime.}\BibitemShut {Stop}%
\bibitem [{GHD({\natexlab{a}})}]{GHD_and_Wigner}%
  \BibitemOpen
  \href@noop {} {}\bibinfo {howpublished} {In the Tonks-Girardeau limit, there
  is an exact equivalence between GHD and the semiclassical Wigner function
  model of a free Fermi gas \cite{quripples,fateofqushock,largescale_Doyon}.
  This model predicts a smooth density profile of the semiclassical shock,
  whereas quantum corrections, derived in
  Refs.~\cite{universalfermi,quripples,protopopov}, predict small-amplitude
  oscillations (`density ripples') on top of the smooth density profile. These
  oscillations persist within a finite extent of the shock front for times well
  beyond shock formation \cite{fateofqushock}, and have a frequency chirp that
  is in the opposite direction to the chirp of the interference fringes seen in
  GPE simulations of dispersive shock waves in the weakly interacting regime.}
  ({\natexlab{a}})\BibitemShut {NoStop}%
\end{thebibliography}

%


~

\newpage

 \setlength{\voffset}{-1.7cm}
 \setlength{\hoffset}{-1.8cm}

\begin{figure}
\includegraphics{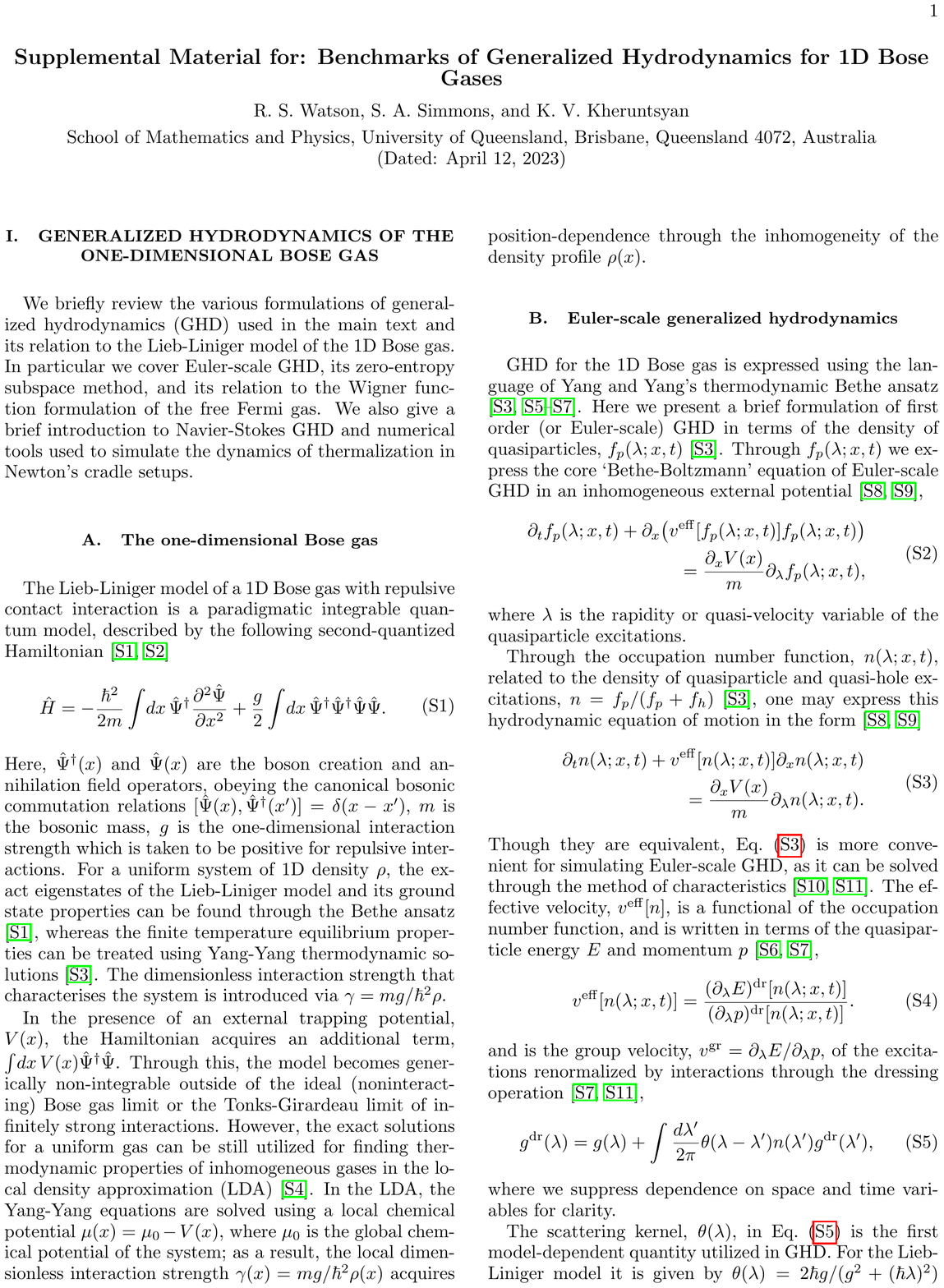}
\end{figure}

\begin{figure}
\includegraphics{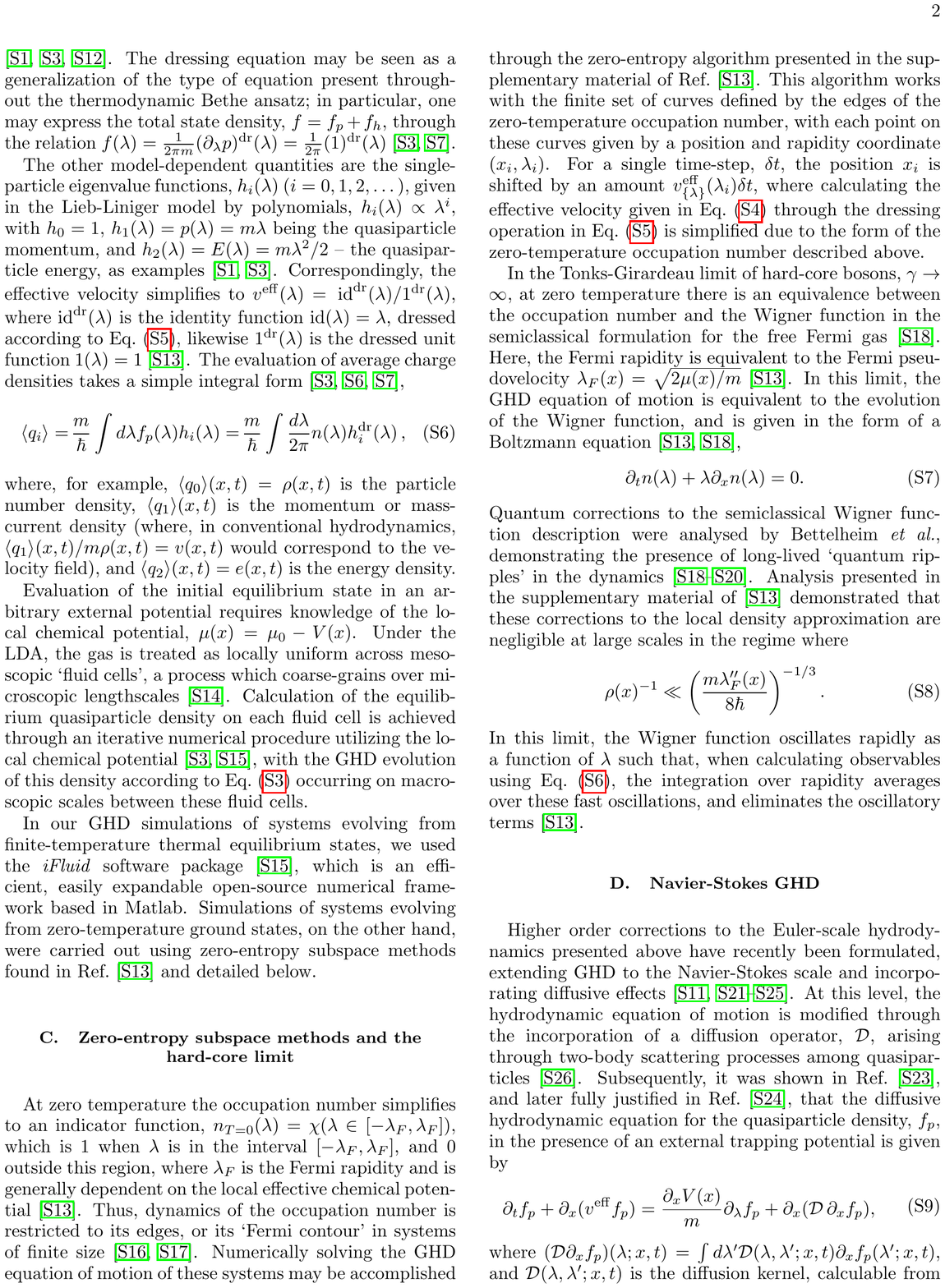}
\end{figure}

\begin{figure}
\includegraphics{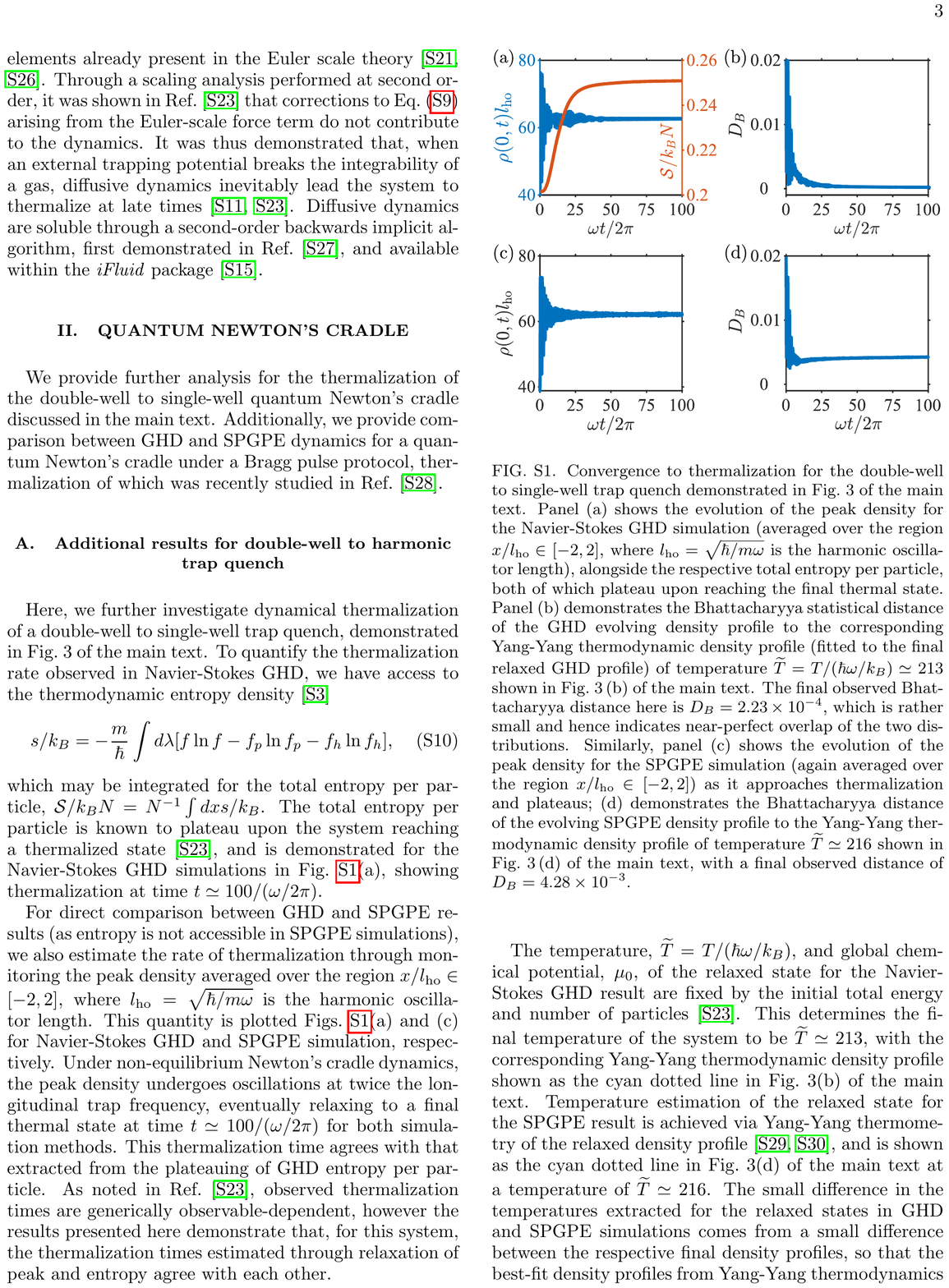}
\end{figure}

\begin{figure}
\includegraphics{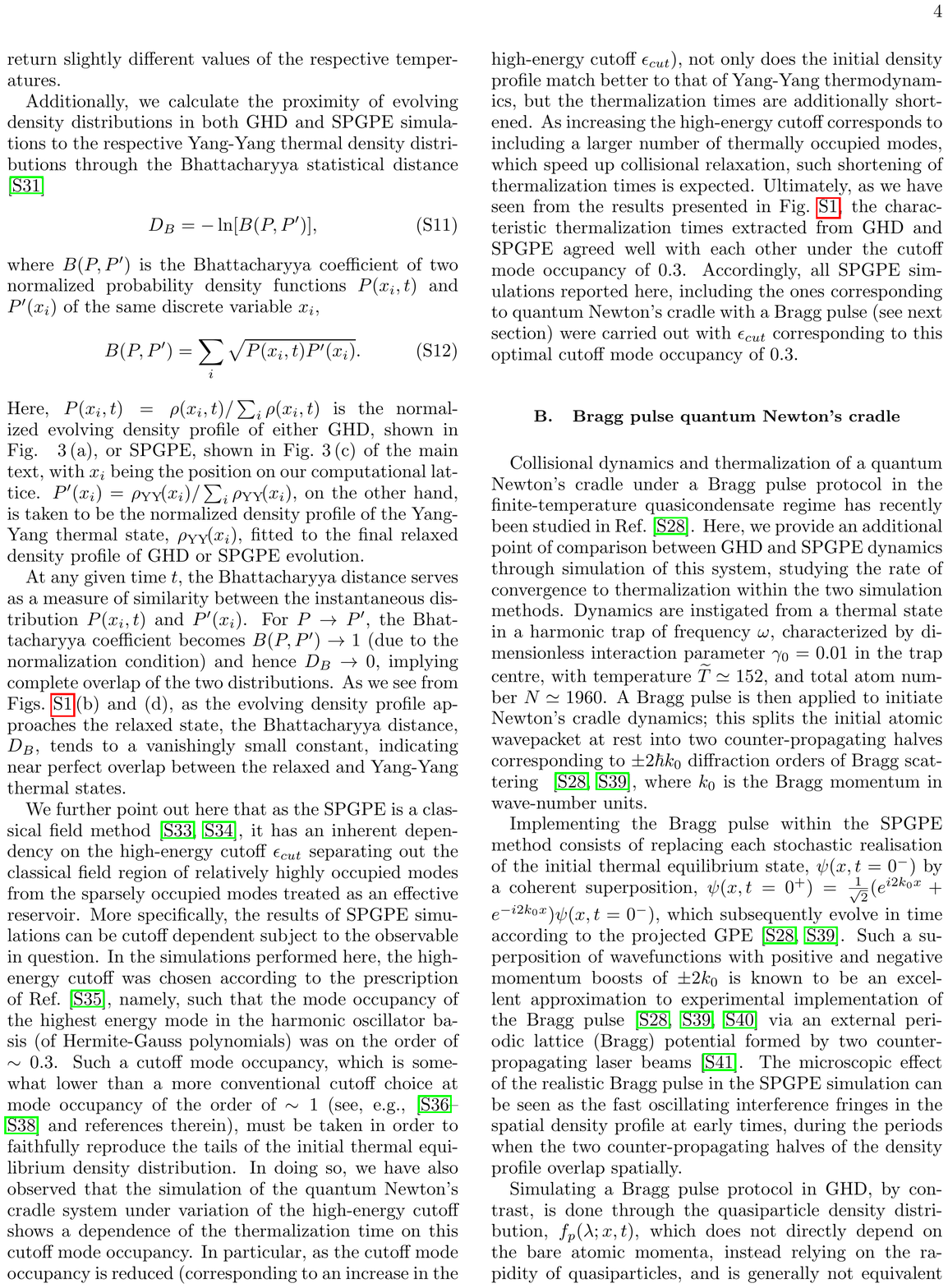}
\end{figure}

\begin{figure}
\includegraphics{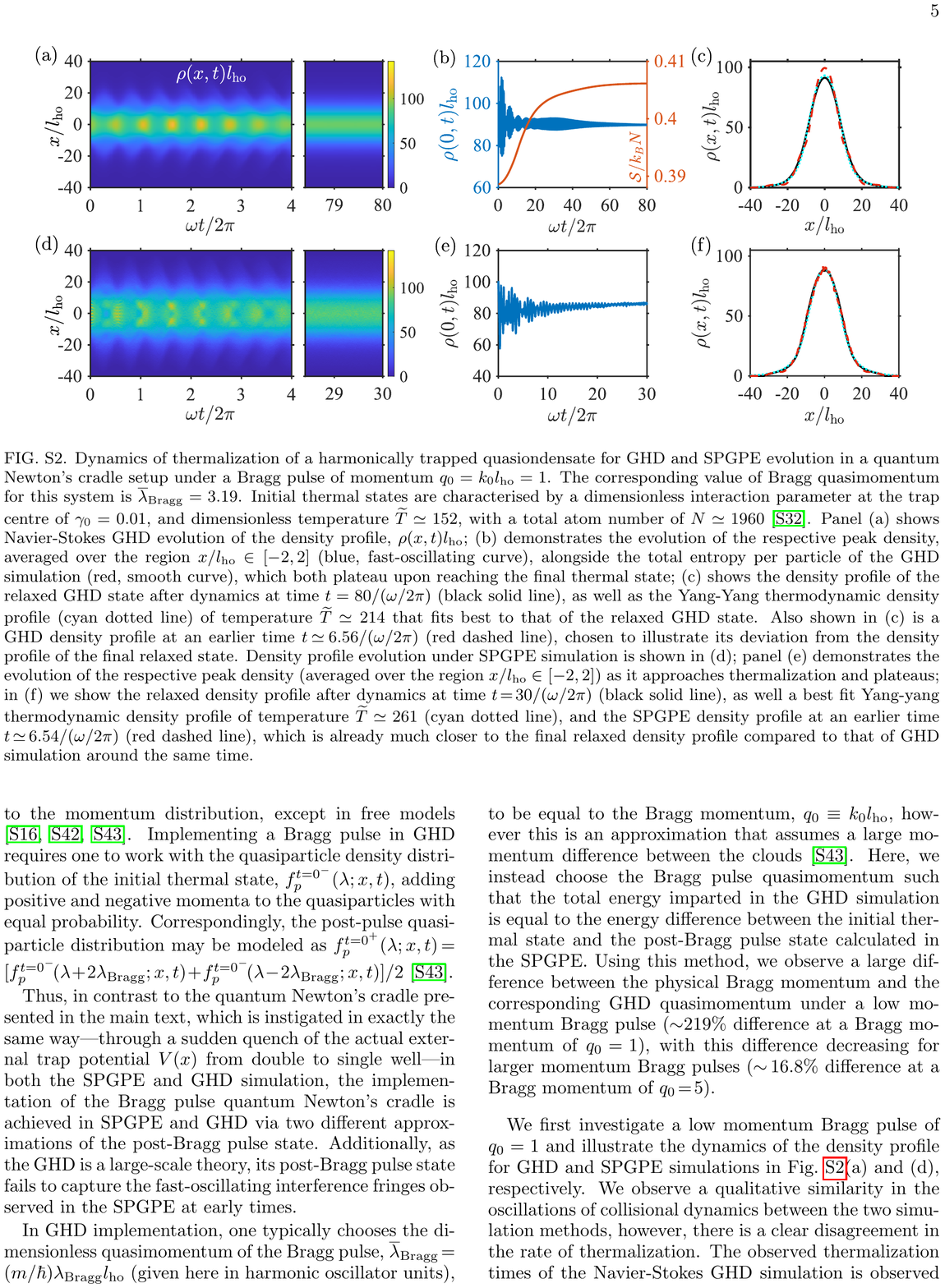}
\end{figure}

\begin{figure}
\includegraphics{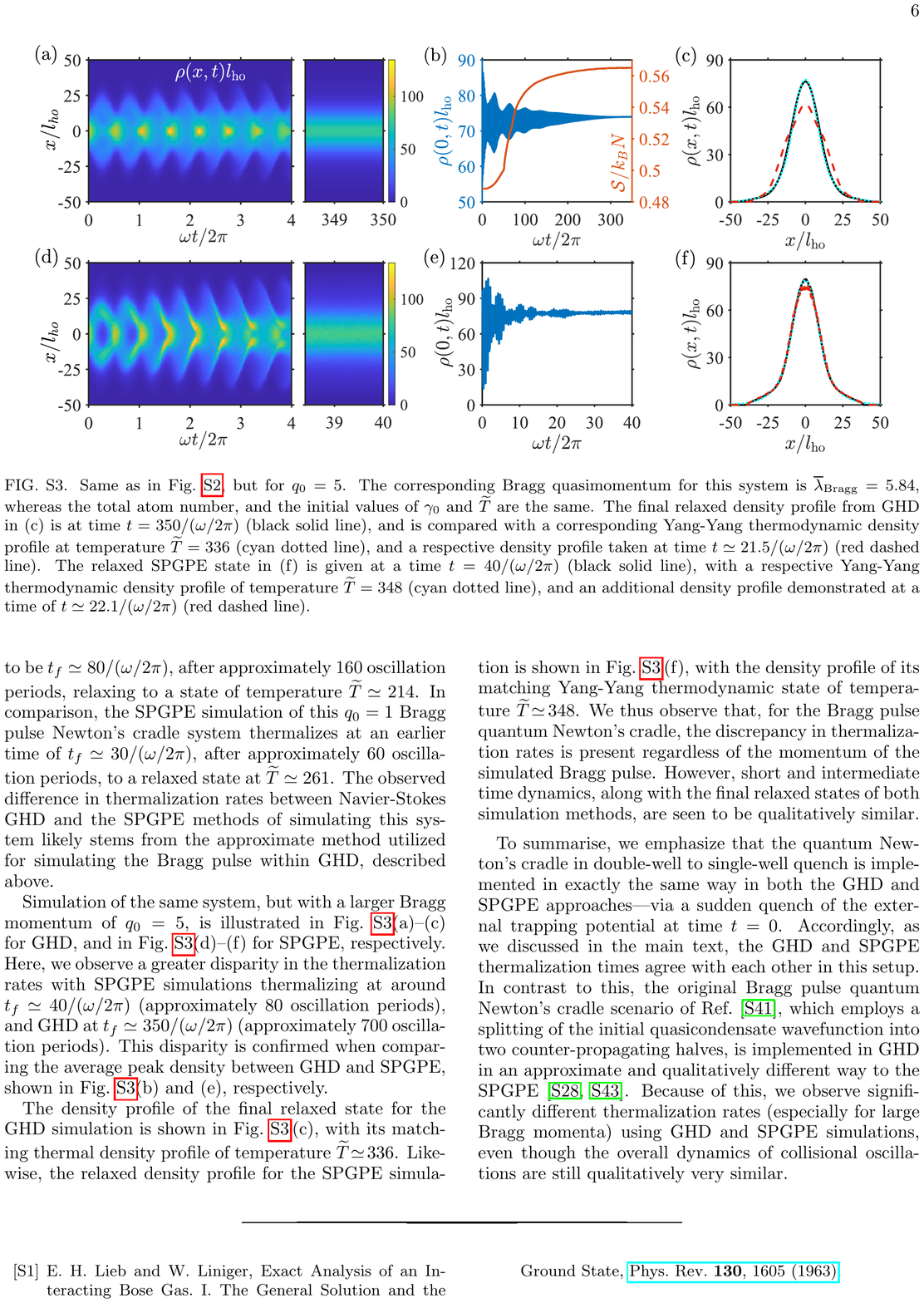}
\end{figure}

\begin{figure}
\includegraphics{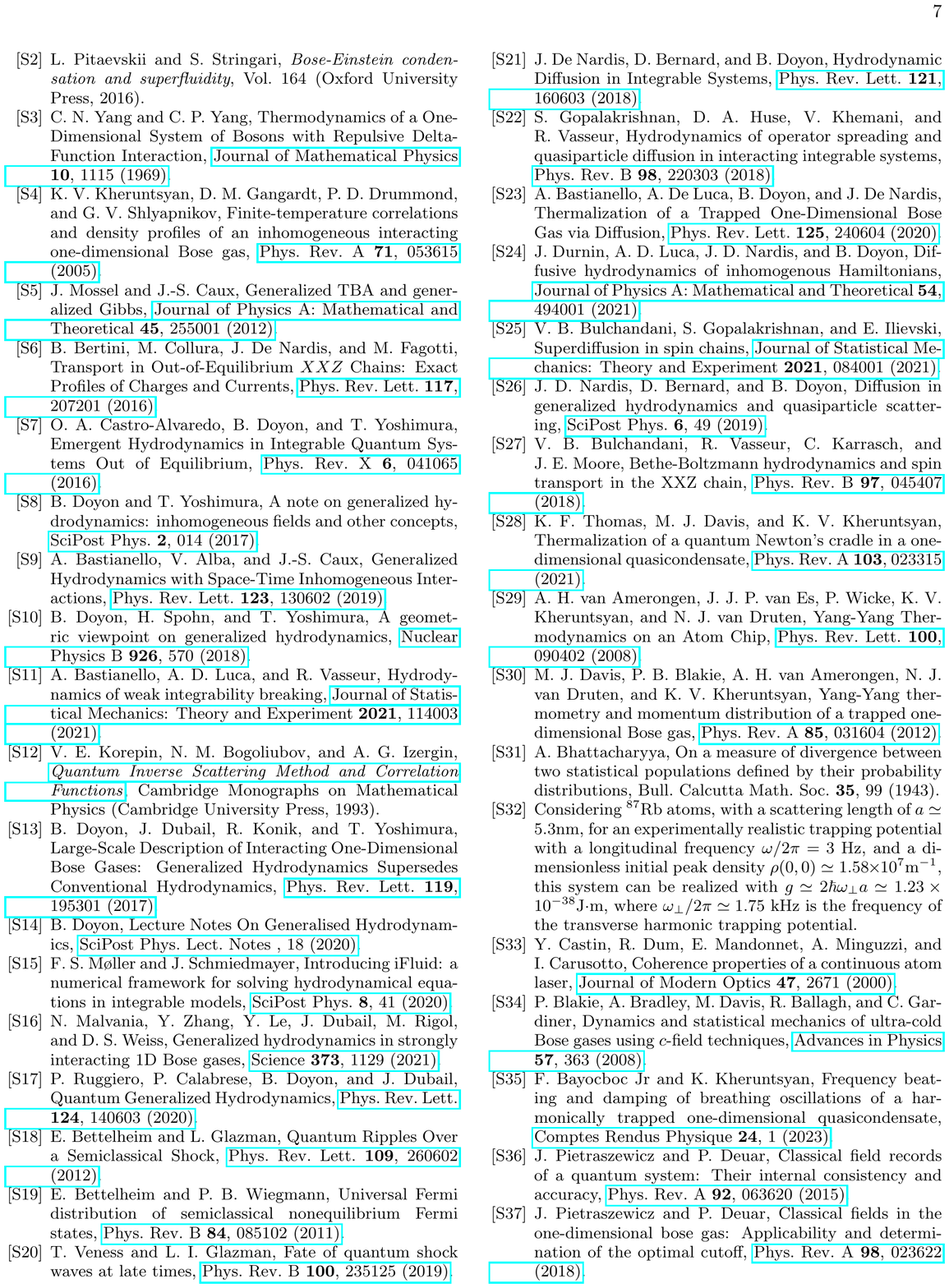}
\end{figure}

\begin{figure}
\includegraphics{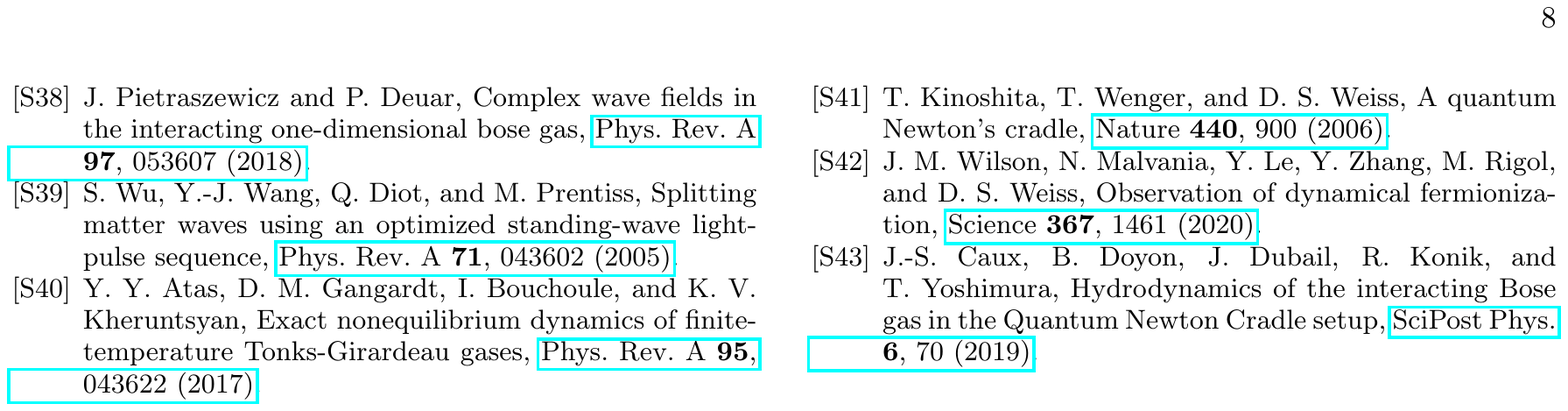}
\end{figure}

\end{document}


\flushbottom

\onecolumngrid
\newpage
\begin{center}
  \textbf{\large Supplemental Material for: Benchmarks of Generalized Hydrodynamics for 1D Bose Gases}\\[.2cm]
  R. S. Watson, S. A. Simmons, and K. V. Kheruntsyan \\[.1cm]
  {School of Mathematics and Physics, University of Queensland, Brisbane,
  Queensland 4072, Australia\\}
(Dated: \today)\\[1cm]
\end{center}

\twocolumngrid

\setcounter{equation}{0}
\setcounter{figure}{0}
\setcounter{table}{0}
\setcounter{page}{1}
\renewcommand{\theequation}{S\arabic{equation}}
\renewcommand{\thefigure}{S\arabic{figure}}
\renewcommand{\bibnumfmt}[1]{[S#1]}
\renewcommand{\citenumfont}[1]{S#1}

\section{Generalized Hydrodynamics of the one-dimensional Bose Gas}\label{sec:GHD_Appendix}

We briefly review the various formulations of generalized hydrodynamics (GHD) used in the main text and its relation to the Lieb-Liniger model of the 1D Bose gas. In particular we cover Euler-scale GHD, its zero-entropy subspace method, and its relation to the Wigner function formulation of the free Fermi gas. We also give a brief introduction to Navier-Stokes GHD and numerical tools used to simulate the dynamics of thermalization in Newton's cradle setups.

\subsection{The one-dimensional Bose gas}

The Lieb-Liniger model of a 1D Bose gas with  repulsive contact interaction is a paradigmatic integrable quantum model, described by the following second-quantized Hamiltonian \cite{liebliniger,pitaevskii2016bose}
\begin{equation}{\label{eq:LL_Ham}}
    \begin{split}
        \hat{H} =- \frac{\hbar^2}{2m} \int \!dx\, \hat{\Psi}^\dagger \frac{\partial^2 \hat{\Psi}}{\partial x^2} 
        + \frac{g}{2} \int \!dx\, \hat{\Psi}^\dagger \hat{\Psi}^\dagger \hat{\Psi} \hat{\Psi}.
    \end{split}
\end{equation}
Here, $\hat{\Psi}^\dagger(x)$ and $\hat{\Psi}(x)$ are the boson creation and annihilation field operators, obeying the canonical bosonic commutation relations $[\hat{\Psi}(x),\hat{\Psi}^\dagger(x')]=\delta(x - x')$, $m$ is the bosonic mass, $g$ is the one-dimensional interaction strength which is taken to be positive for repulsive interactions.
For a uniform system  of 1D density $\rho$, the exact eigenstates of the Lieb-Liniger model and its ground state properties can be found through the Bethe ansatz \cite{liebliniger}, whereas the finite temperature equilibrium  properties can be treated using Yang-Yang  thermodynamic solutions \cite{yang1969thermodynamics}. 
The dimensionless interaction strength that characterises the system is introduced via $\gamma = m g / \hbar^2 \rho$.

In the presence of an external trapping potential, $V(x)$, the Hamiltonian acquires an additional term, $\int\! dx\, V(x) \hat{\Psi}^\dagger \hat{\Psi}$. Through this, the model becomes generically non-integrable outside of the ideal (noninteracting) Bose gas limit or the Tonks-Girardeau limit of infinitely strong interactions. However, the exact solutions for a uniform gas can be still utilized for finding thermodynamic properties of inhomogeneous gases in the local density approximation (LDA) \cite{kheruntsyan2005finite}. In the LDA, the Yang-Yang equations are solved using  a local chemical potential $\mu(x)=\mu_0-V(x)$, where $\mu_0$ is the global chemical potential of the system; as a result, the local dimensionless interaction strength $\gamma(x)=mg/\hbar^2\rho(x)$ acquires position-dependence through the inhomogeneity of the density profile $\rho(x)$.

\subsection{Euler-scale generalized hydrodynamics}

GHD for the 1D Bose gas is expressed using the language of Yang and Yang's thermodynamic Bethe ansatz \cite{yang1969thermodynamics,mossel2012generalized,Bertini_2016_Transport,Castro-Alvaredo_2016_Emergent}. Here we present a brief formulation of first order (or Euler-scale) GHD in terms of the density of quasiparticles, $f_p(\lambda;x,t)$ \cite{yang1969thermodynamics}. Through $f_p(\lambda;x,t)$ we express the core `Bethe-Boltzmann' equation of Euler-scale GHD in an inhomogeneous external potential \cite{doyon2017note,bastianello2019generalized},
\begin{equation}
\begin{split}
    \partial_t f_p(\lambda;x,t) + \partial_x \big( v^\mathrm{eff}[f_p(\lambda;x,t)] f_p(\lambda;x,t) \big) \\
    =    \frac{\partial_x V(x)}{m}  \partial_\lambda f_p(\lambda;x,t) ,
\end{split}
\end{equation}
where $\lambda$ is the rapidity or quasi-velocity variable of the quasiparticle excitations.

Through the occupation number function, $n(\lambda;x,t)$, related to the density of quasiparticle and quasi-hole excitations, $n = f_p/(f_p + f_h)$ \cite{yang1969thermodynamics}, one may express this hydrodynamic equation of motion in the form \cite{doyon2017note,bastianello2019generalized}
\begin{equation}\label{eq:ghd_betheboltzmann_occNum}
\begin{split}
    \partial_t n(\lambda;x,t) + v^\mathrm{eff}[n(\lambda;x,t)] \partial_x n(\lambda;x,t)\\ = \frac{\partial_x V(x)}{m} \partial_\lambda n(\lambda;x,t).
    \end{split}
\end{equation}
Though they are equivalent, Eq.~\eqref{eq:ghd_betheboltzmann_occNum} is more convenient for simulating Euler-scale GHD, as it can be solved through the method of characteristics \cite{doyon2018geometric,bastianello2021hydrodynamics}.  The effective velocity, $v^\mathrm{eff}[n]$, is a functional of the occupation number function, and is written in terms of the quasiparticle energy $E$ and momentum $p$ \cite{Castro-Alvaredo_2016_Emergent,Bertini_2016_Transport},
\begin{equation}\label{eq:effective_velocity}
v^\mathrm{eff}[n(\lambda;x,t)] = \frac{(\partial_\lambda E)^\mathrm{dr}[n(\lambda;x,t)]}{(\partial_\lambda p)^\mathrm{dr}[n(\lambda;x,t)]}.
\end{equation}
and is the group velocity, $v^\mathrm{gr} = \partial_\lambda E / \partial_\lambda p$, of the excitations renormalized by interactions through the dressing operation \cite{bastianello2021hydrodynamics,Castro-Alvaredo_2016_Emergent},
\begin{equation}\label{eq:dressing}
    g^\mathrm{dr}(\lambda) = g(\lambda) + \int \frac{d \lambda'}{2 \pi} \theta(\lambda - \lambda') n(\lambda') g^\mathrm{dr}(\lambda'),
\end{equation}
where we suppress dependence on space and time variables for clarity. 

The scattering kernel, $\theta(\lambda)$, in Eq.~\eqref{eq:dressing} is the first model-dependent quantity utilized in GHD. For the Lieb-Liniger model it is given by $\theta(\lambda) = 2 \hbar g / (g^2 + (\hbar \lambda)^2)$ \cite{liebliniger,yang1969thermodynamics,korepin1993}. 
The dressing equation may be seen as a generalization of the type of equation present throughout the thermodynamic Bethe ansatz; in particular, one may express the total state density, $f = f_p + f_h$, through the relation $f(\lambda) = \frac{1}{2 \pi m}(\partial_\lambda p)^\mathrm{dr}(\lambda) = \frac{1}{2 \pi}(1)^\mathrm{dr}(\lambda)$ \cite{yang1969thermodynamics,Castro-Alvaredo_2016_Emergent}. 

The other model-dependent quantities are the single-particle eigenvalue functions, $h_i(\lambda)$ ($i = 0, 1, 2, \dots$), given in the Lieb-Liniger model by polynomials, $h_i(\lambda) \propto \lambda^i$, with $h_0=1$, $h_1(\lambda) = p(\lambda) = m \lambda$ being the quasiparticle momentum, and $h_2(\lambda) = E(\lambda) = m \lambda^2 / 2 $ -- the quasiparticle energy, as examples \cite{liebliniger,yang1969thermodynamics}. Correspondingly, the effective velocity simplifies to $v^\mathrm{eff}(\lambda) =  \, \mathrm{id}^\mathrm{dr} (\lambda) / 1^\mathrm{dr} (\lambda)$, where $\mathrm{id}^{\mathrm{dr}}(\lambda)$ is the identity function $\mathrm{id}(\lambda) = \lambda$, dressed according to Eq.~\eqref{eq:dressing}, likewise $1^{\mathrm{dr}}(\lambda)$ is the dressed unit function $1(\lambda) = 1$ \cite{largescale_Doyon}.
The evaluation of average charge densities takes a simple integral form \cite{yang1969thermodynamics,Castro-Alvaredo_2016_Emergent,Bertini_2016_Transport},
\begin{equation}\label{eq:observables}
    \langle q_i \rangle =\! \frac{m}{\hbar} \int d \lambda f_p(\lambda) h_i(\lambda) = \! \frac{m}{\hbar} \int \frac{d \lambda}{2 \pi} n(\lambda) h^\mathrm{dr}_i(\lambda) \, , 
\end{equation}
where, for example, $\langle q_0 \rangle (x,t) = \rho(x,t)$ is the particle number density, $\langle q_1 \rangle (x,t)$ is the momentum or mass-current density (where, in conventional hydrodynamics, $\langle q_1 \rangle (x,t) / m \rho(x,t) =v(x,t)$ would correspond to the velocity field), and $\langle q_2 \rangle (x,t) = e(x,t)$ is the energy density.

Evaluation of the initial equilibrium state in an arbitrary external potential requires knowledge of the local  chemical potential, $\mu(x) = \mu_0 - V(x)$. Under the LDA, the gas is treated as locally uniform across mesoscopic `fluid cells', a process which coarse-grains over microscopic lengthscales \cite{doyon2019lecture}. Calculation of the equilibrium quasiparticle density on each fluid cell is achieved through an iterative numerical procedure utilizing the local chemical potential \cite{yang1969thermodynamics,moller2020introducing}, with the GHD evolution of this density according to Eq.~\eqref{eq:ghd_betheboltzmann_occNum} occurring on macroscopic scales between these fluid cells.

In our GHD simulations of systems evolving from finite-temperature thermal equilibrium states, we used the \emph{iFluid}  software package \cite{moller2020introducing}, which is an efficient, easily expandable open-source numerical framework based in Matlab. Simulations of systems evolving from zero-temperature ground states, on the other hand, were carried out using zero-entropy subspace methods found in Ref.~\cite{largescale_Doyon} and detailed below.

\subsection{Zero-entropy subspace methods and the hard-core limit}

At zero temperature the occupation number simplifies to an indicator function, $n_{T=0}(\lambda) = \chi(\lambda \in [- \lambda_F, \lambda_F])$, which is $1$ when $\lambda$ is in the interval $[- \lambda_F, \lambda_F]$, and $0$ outside this region, where $\lambda_F$ is the Fermi rapidity and is generally dependent on the local effective chemical potential \cite{largescale_Doyon}. Thus, dynamics of the occupation number is restricted to its edges, or its `Fermi contour' in systems of finite size \cite{malvania2020generalized,ruggiero2020quantum}. Numerically solving the GHD equation of motion of these systems may be accomplished through the zero-entropy algorithm presented in the supplementary material of Ref.~\cite{largescale_Doyon}. This algorithm works with the finite set of curves defined by the edges of the zero-temperature occupation number, with each point on these curves given by a position and rapidity coordinate $(x_i,\lambda_i)$. For a single time-step, $\delta t$, the position $x_i$ is shifted by an amount $v^\mathrm{eff}_{\{ \lambda \}}(\lambda_i) \delta t$, where calculating the effective velocity given in Eq.~\eqref{eq:effective_velocity} through the dressing operation in Eq.~\eqref{eq:dressing} is simplified due to the form of the zero-temperature occupation number described above.

In the Tonks-Girardeau limit of hard-core bosons, $\gamma \to \infty$, at zero temperature there is an equivalence between the occupation number and the Wigner function in the semiclassical formulation for the free Fermi gas \cite{quripples}. Here, the Fermi rapidity is equivalent to the Fermi pseudovelocity  $\lambda_F(x) =  \sqrt{ 2 \mu(x) / m}$ \cite{largescale_Doyon}. In this limit, the GHD equation of motion is equivalent to the evolution of the Wigner function, and is given in the form of a Boltzmann equation \cite{quripples,largescale_Doyon},
\begin{equation}\label{eq:Boltzmann}
    \partial_t n(\lambda) + \lambda \partial_x n(\lambda) = 0.
\end{equation}
Quantum corrections to the semiclassical Wigner function description were analysed by Bettelheim \textit{et al.}, demonstrating the presence of long-lived `quantum ripples' in the dynamics \cite{universalfermi,quripples,fateofqushock}. Analysis presented in the supplementary material of \cite{largescale_Doyon} demonstrated that these corrections to the local density approximation are negligible at large scales in the regime where
\begin{equation}
    \rho(x)^{-1} \ll \left(\frac{m \lambda_F''(x)}{8 \hbar} \right)^{-1/3}.
\end{equation}
In this limit, the Wigner function oscillates rapidly as a function of $\lambda$ such that, when calculating observables using Eq.~\eqref{eq:observables}, the integration over rapidity averages over these fast oscillations, and eliminates the oscillatory terms \cite{largescale_Doyon}.

\subsection{Navier-Stokes GHD}

Higher order corrections to the Euler-scale hydrodynamics presented above have recently been formulated, extending GHD to the Navier-Stokes scale and incorporating diffusive effects \cite{DeNardis_Diffusion_2018,gopalakrishnan2018hydrodynamics,bastianello_thermalization_2020,durnin2021diffusive,bastianello2021hydrodynamics,Bulchandani_2021}. At this level, the hydrodynamic equation of motion is modified through the incorporation of a diffusion operator, $\mathcal{D}$, arising through two-body scattering processes among quasiparticles \cite{de2019diffusion}. Subsequently, it was shown in Ref.~\cite{bastianello_thermalization_2020}, and later fully justified in Ref.~\cite{durnin2021diffusive}, that the diffusive hydrodynamic equation for the quasiparticle density, $f_p$, in the presence of an external trapping potential is given by
\begin{equation}\label{eq:DiffusiveGHD}
\begin{split}
\begin{aligned}
\partial_t & f_p + \partial_x(v^\mathrm{eff} f_p) = \frac{\partial_x V(x)}{m} \partial_\lambda f_p + \partial_x(\mathcal{D} \, \partial_x f_p),
\end{aligned}
\end{split}
\end{equation}
where $(\mathcal{D} \partial_x f_p)(\lambda;x,t) = \int d\lambda' \mathcal{D}(\lambda,\lambda'; x,t) \partial_x f_p(\lambda';x,t)$, and $\mathcal{D}(\lambda,\lambda';x,t)$ is the diffusion kernel, calculable from elements already present in the Euler scale theory \cite{DeNardis_Diffusion_2018,de2019diffusion}. 
Through a scaling analysis performed at second order, it was shown in Ref.~\cite{bastianello_thermalization_2020} that corrections to Eq.~\eqref{eq:DiffusiveGHD} arising from the Euler-scale force term do not contribute to the dynamics. It was thus demonstrated that, when an external trapping potential breaks the integrability of a gas, diffusive dynamics inevitably lead the system to thermalize at late times \cite{bastianello_thermalization_2020,bastianello2021hydrodynamics}. Diffusive dynamics are soluble through a second-order backwards implicit algorithm, first demonstrated in Ref.~\cite{bulchandani2018bethe}, and available within the \textit{iFluid} package \cite{moller2020introducing}.

\section{Quantum Newton's cradle}
We provide further analysis for the thermalization of the double-well to single-well quantum Newton's cradle discussed in the main text. Additionally, we provide comparison between GHD and SPGPE dynamics for a quantum Newton's cradle under a Bragg pulse protocol, thermalization of which was recently studied in Ref.~\cite{kieran_paper}.

\subsection{Additional results for double-well to harmonic trap quench}

\begin{figure}[tb]
    \includegraphics[scale=0.67]{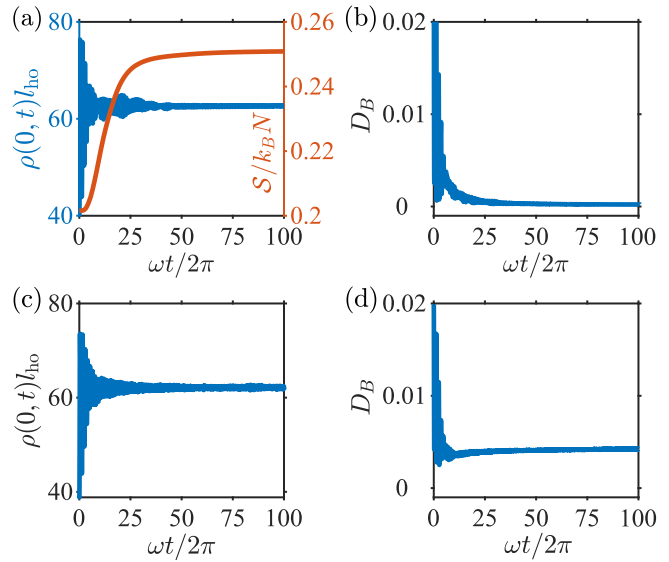}
    \caption{Convergence to thermalization for the double-well to single-well trap quench demonstrated in Fig.~3 of the main text. Panel (a) shows the evolution of the peak density for the Navier-Stokes GHD simulation (averaged over the region $x/l_\mathrm{ho} \in [-2,2]$, where $l_\mathrm{ho} = \sqrt{\hbar / m \omega}$ is the harmonic oscillator length), alongside the respective total entropy per particle, both of which plateau upon reaching the final thermal state. Panel (b) demonstrates the Bhattacharyya statistical distance of the GHD evolving density profile to the corresponding Yang-Yang thermodynamic density profile (fitted to the final relaxed GHD profile) of temperature $\widetilde{T} = T/(\hbar \omega /k_B)\simeq213$ shown in Fig.~3\,(b) of the main text. The final observed Bhattacharyya distance here is $D_B = 2.23\times10^{-4}$, which is rather small and hence indicates near-perfect overlap of the two distributions. Similarly, panel (c) shows the evolution of the peak density for the SPGPE simulation (again averaged over the region $x/l_\mathrm{ho} \in [-2,2]$) as it approaches thermalization and plateaus; (d) demonstrates the Bhattacharyya distance of the evolving SPGPE density profile to the Yang-Yang thermodynamic density profile of temperature $\widetilde{T}\simeq216$ shown in Fig.~3\,(d) of the main text, with a final observed distance of $D_B = 4.28\times10^{-3}$.
   }
    \label{fig:Therm_SymDoubleWell}
\end{figure}

Here, we further investigate dynamical thermalization of a double-well to single-well trap quench, demonstrated in Fig.~3 of the main text. To quantify the thermalization rate observed in Navier-Stokes GHD, we have access to the thermodynamic entropy density \cite{yang1969thermodynamics}
\begin{equation}
    s/k_B = -\frac{m}{\hbar}\int d\lambda [ f \ln f - f_p \ln f_p - f_h \ln f_h ],
\end{equation}
which may be integrated for the total entropy per particle, $\mathcal{S}/k_B N = N^{-1} \int dx s/k_B$. The total entropy per particle is known to plateau upon the system reaching a thermalized state \cite{bastianello_thermalization_2020}, and is demonstrated for the Navier-Stokes GHD simulations in Fig.~\ref{fig:Therm_SymDoubleWell}(a), showing thermalization at time $t \simeq 100/(\omega/2\pi)$.

For direct comparison between GHD and SPGPE results (as entropy is not accessible in SPGPE simulations), we also estimate the rate of thermalization through monitoring the peak density averaged over the region $x/l_\mathrm{ho} \in [-2,2]$, where $l_\mathrm{ho} = \sqrt{\hbar / m \omega}$ is the harmonic oscillator length. This quantity is plotted Figs.~\ref{fig:Therm_SymDoubleWell}(a) and (c) for Navier-Stokes GHD and SPGPE simulation, respectively. Under non-equilibrium Newton's cradle dynamics, the peak density undergoes oscillations at twice the longitudinal trap frequency, eventually relaxing to a final thermal state at time $t \simeq 100/(\omega/2\pi)$ for both simulation methods. This thermalization time agrees with that extracted from the plateauing of GHD entropy per particle. As noted in Ref.~\cite{bastianello_thermalization_2020}, observed thermalization times are generically observable-dependent, however the results presented here demonstrate that, for this system, the thermalization times estimated through relaxation of peak and entropy agree with each other.

The temperature, $\widetilde{T} = T/(\hbar \omega /k_B)$, and global chemical potential, $\mu_0$, of the relaxed state for the Navier-Stokes GHD result are fixed by the initial total energy and number of particles \cite{bastianello_thermalization_2020}. This determines the final temperature of the system to be $\widetilde{T} \simeq 213$, with the corresponding Yang-Yang thermodynamic density profile shown as the cyan dotted line in Fig.~3(b) of the main text. Temperature estimation of the relaxed state for the SPGPE result is achieved via Yang-Yang thermometry of the relaxed density profile \cite{Karen_Yang_2008,davis2012yang}, and is shown as the cyan dotted line in Fig.~3(d) of the main text at a temperature of $\widetilde{T} \simeq 216$. The small difference in the temperatures extracted for the relaxed states in GHD and SPGPE simulations comes from a small difference between the respective final density profiles, so that the best-fit density profiles from Yang-Yang thermodynamics return slightly different values of the respective temperatures. 

Additionally, we calculate the proximity of evolving density distributions in both GHD and SPGPE simulations to the respective Yang-Yang thermal density distributions   through the Bhattacharyya statistical distance \cite{bhattacharyya1943measure}
\begin{equation}
    D_B = -\ln [ B(P,P') ],
\end{equation}
where $B(P,P')$ is the Bhattacharyya coefficient of two normalized probability density functions $P(x_i,t)$ and $P'(x_i)$ of the same discrete variable $x_i$, 
\begin{equation}
    B(P,P') = \sum_{i} \sqrt{P(x_i,t) P'(x_i)}.
\end{equation}
Here, $P(x_i,t) = \rho(x_i,t)/\sum_{i} \rho(x_i,t)$ is the normalized evolving density profile of either GHD, shown in Fig.~
3\,(a), or SPGPE, shown in Fig.~3\,(c) of the main text, with $x_i$ being the position on our computational lattice. $P'(x_i) = \rho_\mathrm{YY}\!(x_i)/\sum_{i}\rho_\mathrm{YY}\!(x_i)$, on the other hand, is taken to be the normalized density profile of the Yang-Yang thermal state, $\rho_\mathrm{YY}\!(x_i)$, fitted to the final relaxed density profile of GHD or SPGPE evolution. 

At any given time $t$, the Bhattacharyya distance serves as a measure of similarity between the instantaneous distribution $P(x_i,t)$ and $P'(x_i)$. For $P\to P'$, the Bhattacharyya coefficient becomes $B(P,P')\to 1$ (due to the normalization condition) and hence $D_B\to 0$, implying complete overlap of the two distributions. As we see from Figs.~\ref{fig:Therm_SymDoubleWell}\,(b) and (d), as the evolving density profile approaches the relaxed state, the Bhattacharyya distance, $D_B$, tends to a vanishingly small constant, indicating near perfect overlap   between the relaxed and Yang-Yang thermal states.

\begin{figure*}[tbp]
    \includegraphics[width=16.7cm]{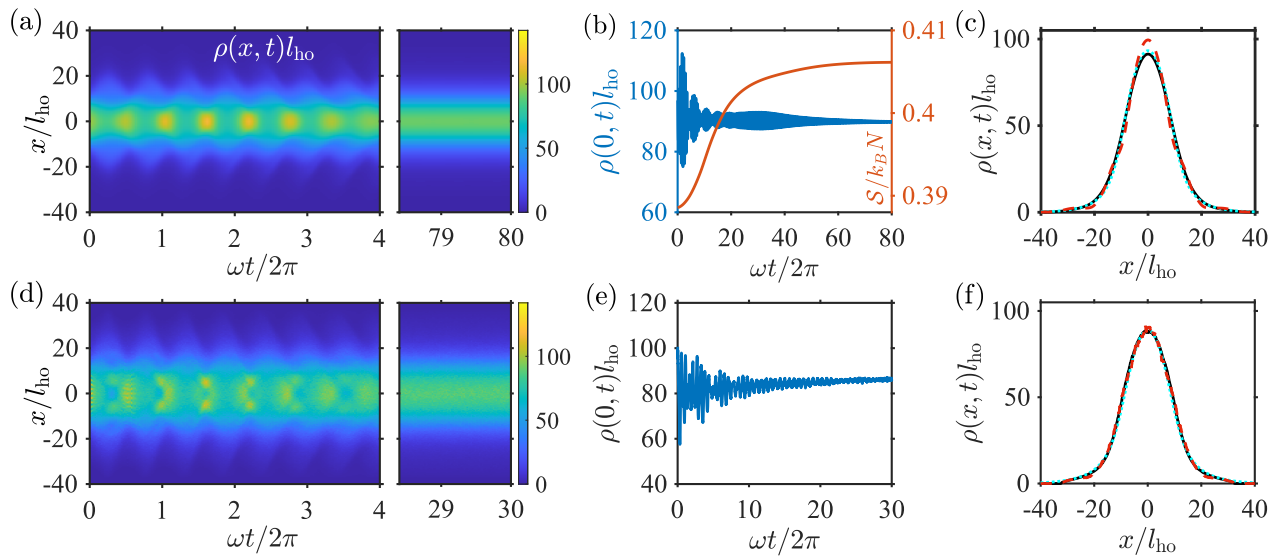}
    \caption{Dynamics of thermalization of a harmonically trapped quasiondensate for GHD and SPGPE evolution in a quantum Newton's cradle setup under a Bragg pulse of momentum $q_0 = k_0 l_\mathrm{ho} = 1$. The corresponding value of Bragg quasimomentum for this system is $\overline{\lambda}_{\mathrm{Bragg}} = 3.19$. Initial thermal states are characterised by a dimensionless interaction parameter at the trap centre of $\gamma_0 = 0.01$, and dimensionless temperature $\widetilde{T}\simeq 152$, with a total atom number of $N\simeq1960$ \cite{BraggPulseNewtonsCradle_Parameters}.
    Panel (a) shows Navier-Stokes GHD evolution of the density profile, $\rho(x,t) l_\mathrm{ho}$; (b) demonstrates the evolution of the respective peak density, averaged over the region $x/l_\mathrm{ho} \in [-2,2]$ (blue, fast-oscillating curve), alongside the total entropy per particle of the GHD simulation (red, smooth curve), which both plateau upon reaching the final thermal state; (c) shows the density profile of the relaxed GHD state after dynamics at time $t = 80/(\omega/2\pi)$ (black solid line), as well as the Yang-Yang thermodynamic density profile (cyan dotted line) of temperature $\widetilde{T} \simeq 214$ that fits best to that of the relaxed GHD state. Also shown in (c) is a GHD density profile at an earlier time $t \!\simeq\! 6.56 / (\omega / 2 \pi)$ (red dashed line), chosen to illustrate its deviation from the density profile of the final relaxed state. Density profile evolution under SPGPE simulation is shown in (d); panel (e) demonstrates the evolution of the respective peak density (averaged over the region $x/l_\mathrm{ho} \in [-2,2]$) as it approaches thermalization and plateaus; in (f) we show the relaxed density profile after dynamics at time $t \!=\! 30/(\omega/2\pi)$ (black solid line), as well a best fit Yang-yang thermodynamic density profile of temperature $\widetilde{T} \simeq 261$ (cyan dotted line), and the SPGPE density profile at an earlier time $t\! \simeq\! 6.54 / (\omega / 2 \pi)$ (red dashed line), which is already much closer to the final relaxed density profile compared to that of GHD simulation around the same time.}
\label{fig:q0_1}
\end{figure*}

We further point out here that as the SPGPE is a classical field method \cite{castin2000,Blakie_cfield_2008}, it has an inherent dependency on the high-energy cutoff $\epsilon_{cut}$ separating out the classical field region of relatively highly occupied modes from the sparsely occupied modes treated as an effective reservoir. More specifically, the results of SPGPE simulations can be cutoff dependent subject to the observable in question. In the simulations performed here, the high-energy cutoff was chosen according to the prescription of Ref.~\cite{bayocboc2022frequency}, namely, such that the mode occupancy of the highest energy mode in the harmonic oscillator basis (of Hermite-Gauss polynomials) was on the order of $\sim 0.3$. Such a cutoff mode occupancy, which is somewhat lower than a more conventional cutoff choice at mode occupancy of the order of $\sim 1$ (see, e.g., \cite{Pietraszeqicz2015classical,Pietraszewicz2018classical,Pietraszewicz2018complex} and references therein), must be taken in order to faithfully reproduce the tails of the initial thermal equilibrium density distribution. In doing so, we have also observed that the simulation of the quantum Newton's cradle system under variation of the high-energy cutoff shows a dependence of the thermalization time on this cutoff mode occupancy. In particular, as the cutoff mode occupancy is reduced (corresponding to an increase in the high-energy cutoff $\epsilon_{cut}$), not only does the initial density profile match better to that of Yang-Yang thermodynamics, but the thermalization times are additionally shortened. As increasing the high-energy cutoff corresponds to including a larger number of thermally occupied modes, which speed up collisional relaxation, such shortening   of thermalization times is expected. Ultimately, as we have seen from the results presented in Fig.~\ref{fig:Therm_SymDoubleWell}, the characteristic thermalization times extracted from GHD and SPGPE agreed well with each other under the cutoff mode occupancy of $0.3$. Accordingly, all SPGPE simulations reported here, including the ones corresponding to quantum Newton's cradle with a Bragg pulse (see next section) were carried out with $\epsilon_{cut}$ corresponding to this optimal cutoff mode occupancy of $0.3$.

\begin{figure*}[tbp]    \includegraphics[width=16.7cm]{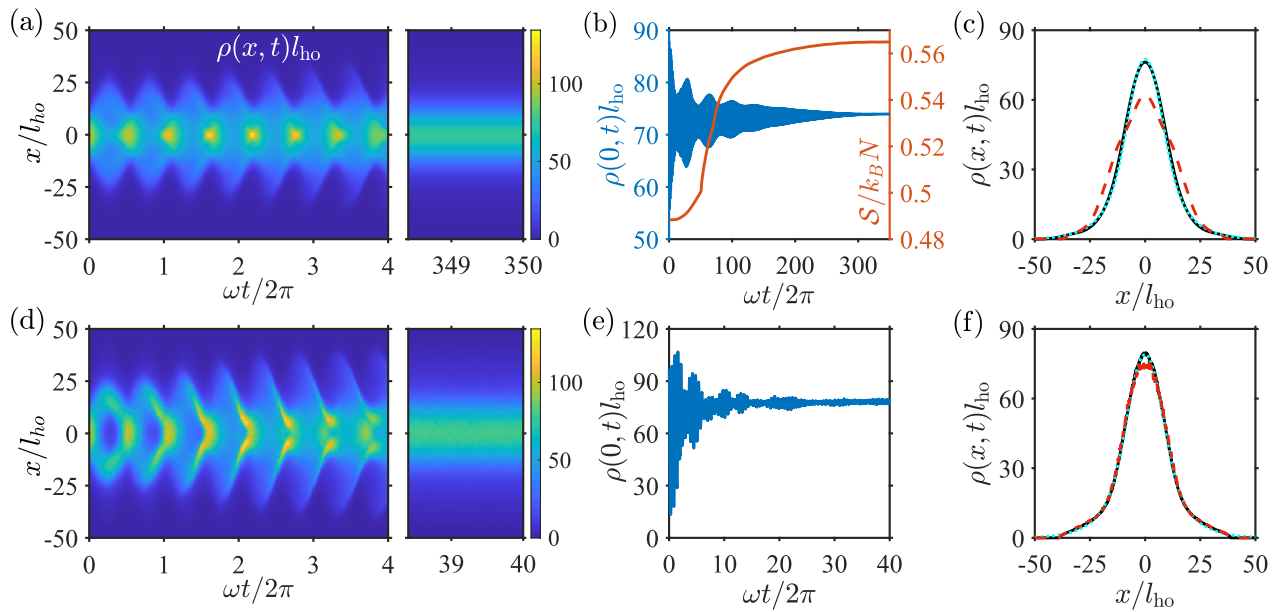}
    \caption{Same as in Fig.~\ref{fig:q0_1}, but for $q_0 = 5$. The corresponding Bragg quasimomentum for this system is $\overline{\lambda}_\mathrm{Bragg} = 5.84$, whereas the total atom number, and the initial values of $\gamma_0$ and $\widetilde{T}$ are the same. The final relaxed density profile from GHD in (c) is at time $t = 350/(\omega/2\pi)$ (black solid line), and is compared with a corresponding Yang-Yang thermodynamic density profile at temperature $\widetilde{T} = 336$ (cyan dotted line), and a respective density profile taken at time $t \simeq 21.5/(\omega/2\pi)$ (red dashed line). The relaxed SPGPE state in (f) is given at a time $t = 40/(\omega/2\pi)$ (black solid line), with a respective Yang-Yang thermodynamic density profile of temperature $\widetilde{T} = 348$ (cyan dotted line), and an additional density profile demonstrated at a time of $t \simeq 22.1/(\omega/2\pi)$ (red dashed line).}
\label{fig:q0_5}
\end{figure*}

\subsection{Bragg pulse quantum Newton's cradle}

Collisional dynamics and thermalization of a quantum Newton's cradle under a Bragg pulse protocol in the finite-temperature quasicondensate regime has recently been studied in Ref.~\cite{kieran_paper}. Here, we provide an additional point of comparison between GHD and SPGPE dynamics through simulation of this system, studying the rate of convergence to thermalization within the two simulation methods. Dynamics are instigated from a thermal state in a harmonic trap of frequency $\omega$, characterized by dimensionless interaction parameter $\gamma_0 = 0.01$ in the trap centre, with temperature $\widetilde{T}\simeq  152$, and total atom number $N\simeq1960$. A Bragg pulse is then applied to initiate Newton's cradle dynamics; this splits the initial atomic wavepacket at rest into two counter-propagating halves corresponding to $\pm 2\hbar k_0$ diffraction orders of Bragg scattering ~\cite{Bragg,kieran_paper}, where $k_0$ is the Bragg momentum in wave-number units.

Implementing the Bragg pulse within the SPGPE method consists of replacing each stochastic realisation of the initial thermal equilibrium state, $\psi(x,t=0^-)$ by a coherent superposition, $\psi(x,t=0^+) = \frac{1}{\sqrt{2}}(e^{i 2 k_0 x} + e^{-i 2 k_0 x}) \psi(x,t=0^-)$, which subsequently evolve in time according to the projected GPE \cite{Bragg,kieran_paper}. Such a superposition of wavefunctions with positive and negative momentum boosts of $\pm2k_0$ is known to be an excellent approximation to experimental implementation of the Bragg pulse \cite{Bragg,kieran_paper,atas_TG_exact} via an external periodic lattice (Bragg) potential formed by two counter-propagating laser beams \cite{kinoshita2006quantum}. The microscopic effect of the realistic Bragg pulse in the SPGPE simulation can be seen as the fast oscillating interference fringes in the spatial density profile at early times, during the periods when the two counter-propagating halves of the density profile overlap spatially.

Simulating a Bragg pulse protocol in GHD, by contrast, is done through the quasiparticle density distribution, $f_p(\lambda;x,t)$, which does not directly depend on the bare atomic momenta, instead relying on the rapidity of quasiparticles, and is generally not equivalent to the momentum distribution, except in free models \cite{Wilson1461, GHD_newtonscradle, malvania2020generalized}. Implementing a Bragg pulse in GHD requires one to work with the quasiparticle density distribution of the initial thermal state, $f^{t=0^-}_p(\lambda;x,t)$, adding positive and negative momenta to the quasiparticles with equal probability. Correspondingly, the post-pulse quasiparticle distribution may be modeled as $f_p^{t=0^+}(\lambda;x,t)\!=\![f^{t=0^-}_p\!(\lambda \!+\! 2 \lambda_\mathrm{Bragg};x,t) \!+\! f^{t =0^-}_p\!(\lambda \!-\! 2\lambda_\mathrm{Bragg};x,t)]/2$ \cite{GHD_newtonscradle}.

Thus, in contrast to the quantum Newton's cradle presented in the main text, which is instigated in exactly the same way---through a sudden quench of the actual external trap potential $V(x)$ from double to single well---in both the SPGPE and GHD simulation, 
the implementation of the Bragg pulse quantum Newton's cradle is achieved in SPGPE and GHD via two different approximations of the post-Bragg pulse state. Additionally, as the GHD is a large-scale theory, its post-Bragg pulse state fails to capture the fast-oscillating interference fringes observed in the SPGPE at early times.

In GHD implementation, one typically chooses the dimensionless quasimomentum of the Bragg pulse, $\overline{\lambda}_\mathrm{Bragg} \!=\! (m/\hbar) \lambda_\mathrm{Bragg} l_\mathrm{ho}$ (given here in  harmonic oscillator units), to be equal to the Bragg momentum, $q_0\equiv k_0 l_{\mathrm{ho}}$, however this is an approximation that assumes a large momentum difference between the clouds \cite{GHD_newtonscradle}. Here, we instead choose the Bragg pulse quasimomentum such that the total energy imparted in the GHD simulation is equal to the energy difference between the initial thermal state and the post-Bragg pulse state calculated in the SPGPE. Using this method, we observe a large difference between the physical Bragg momentum and the corresponding GHD quasimomentum under a low momentum Bragg pulse ($\sim\!\!219 \%$ difference at a Bragg momentum of $q_0 \!=\! 1$), with this difference decreasing for larger momentum Bragg pulses ($\sim\!16.8 \%$ difference at a Bragg momentum of $q_0\! = \!5$).

We first investigate a low momentum Bragg pulse of $q_0 = 1$ and illustrate the dynamics of the density profile for GHD and SPGPE simulations in Fig.~\ref{fig:q0_1}(a) and (d), respectively. We observe a qualitative similarity in the oscillations of collisional dynamics between the two simulation methods, however, there is a clear disagreement in the rate of thermalization. The observed thermalization times of the Navier-Stokes GHD simulation is observed to be $t_f \simeq 80/(\omega/2\pi)$, after approximately 160 oscillation periods, relaxing to a state of temperature $\widetilde{T} \simeq 214$. In comparison, the SPGPE simulation of this $q_0 = 1$ Bragg pulse Newton's cradle system thermalizes at an earlier time of $t_f \simeq 30/(\omega/2\pi)$, after approximately 60 oscillation periods, to a relaxed state at $\widetilde{T} \simeq 261$. The observed difference in thermalization rates between Navier-Stokes GHD and the SPGPE methods of simulating this system likely stems from the approximate method utilized for simulating the Bragg pulse within GHD, described above.

Simulation of the same system, but with a larger Bragg momentum of $q_0 = 5$, is illustrated in Fig.~\ref{fig:q0_5}(a)--(c) for GHD, and in  Fig.~\ref{fig:q0_5}(d)--(f)  for SPGPE, respectively. Here, we observe a greater disparity in the thermalization rates with SPGPE simulations thermalizing at around $t_f \simeq 40/(\omega/2\pi)$ (approximately 80 oscillation periods), and GHD at $t_f \simeq 350/(\omega/2\pi)$ (approximately 700 oscillation periods). This disparity is confirmed when comparing the average peak density between GHD and SPGPE, shown in Fig.~\ref{fig:q0_5}(b) and (e), respectively.

The density profile of the final relaxed state for the GHD simulation is shown in Fig.~\ref{fig:q0_5}\,(c), with its matching thermal density profile of temperature $\widetilde{T} \!\simeq \!336$. Likewise, the relaxed density profile for the SPGPE simulation is shown in Fig.~\ref{fig:q0_5}\,(f), with the density profile of its matching Yang-Yang thermodynamic state of temperature $\widetilde{T} \!\simeq \!348$. We thus observe that, for the Bragg pulse quantum Newton's cradle, the discrepancy in thermalization rates is present regardless of the momentum of the simulated Bragg pulse. However, short and intermediate time dynamics, along with the final relaxed states of both simulation methods, are seen to be qualitatively similar.

To summarise, we emphasize that the quantum Newton's cradle in double-well to single-well quench is implemented in exactly the same way in both the GHD and SPGPE approaches---via a sudden quench of the external trapping potential at time $t=0$. Accordingly, as we discussed in the main text, the GHD and SPGPE thermalization times agree with each other in this setup. In contrast to this, the original Bragg pulse quantum Newton's cradle scenario of Ref.~\cite{kinoshita2006quantum}, which employs a splitting of the initial quasicondensate wavefunction into two counter-propagating halves, is implemented in GHD in an approximate and qualitatively different way to the SPGPE \cite{kieran_paper,GHD_newtonscradle}. Because of this, we observe significantly different thermalization rates (especially for large Bragg momenta) using GHD and SPGPE simulations, even though the overall dynamics of collisional oscillations are still qualitatively very similar. 

\bibliography{man_bib.bib}